\documentclass[iop]{emulateapj-rtx4}
\slugcomment{Accepted for publication in {\it the Astrophysical Journal}}

\newcommand{\fav}{\langle F \rangle}

\begin{document}

\title{PRODUCTION OF THE ${\lowercase{p}}$-PROCESS NUCLEI IN THE
  CARBON-DEFLAGRATION MODEL FOR TYPE I{\lowercase{a}} SUPERNOVAE}

\author{Motohiko Kusakabe\altaffilmark{1}, Nobuyuki
Iwamoto\altaffilmark{2}, and Ken'ichi Nomoto\altaffilmark{3}}

\altaffiltext{1}{Institute for Cosmic Ray Research, University of Tokyo,
  Kashiwa, Chiba 277-8582, Japan\\
{\tt kusakabe@icrr.u-tokyo.ac.jp}}
\altaffiltext{2}{Nuclear Data Center, Japan Atomic Energy Agency, Tokai,
	Ibaraki 319-1195, Japan\\
{\tt iwamoto.nobuyuki@jaea.go.jp}}
\altaffiltext{3}{Institute for the Physics and Mathematics of the
  Universe, University of Tokyo, Kashiwa, Chiba 277-8568, Japan\\
{\tt nomoto@astron.s.u-tokyo.ac.jp}}

\begin{abstract}
We calculate nucleosynthesis of proton-rich isotopes in the carbon-deflagration model for Type~Ia supernovae (SNe~Ia).  The seed
abundances are obtained by calculating the $s$-process nucleosynthesis
that is expected to occur in the repeating helium shell flashes on the
carbon-oxygen (CO) white dwarf (WD) during mass accretion from a binary
companion.  When the deflagration wave passes through the outer layer
of the CO WD, $p$-nuclei are produced by photodisintegration
reactions on $s$-nuclei in a region with the peak temperature ranging
from 1.9 to 3.6 $\times 10^9$~K.  
We confirm the sensitivity of the
$p$-process on the initial distribution of $s$-nuclei.  We show that
the initial C/O ratio in the WD does not much affect the
yield of $p$-nuclei.  On the other hand, the abundance of $^{22}$Ne
left after the $s$-processing has a large influence on the $p$-process
via $^{22}$Ne($\alpha$,$n$) reaction.  
We find that about 50\% of
$p$-nuclides are co-produced when normalized to their solar abundances
in all adopted cases of seed distribution.  Mo and Ru, which are
largely underproduced in Type~II supernovae (SNe~II), are produced
more than in SNe II although they are
underproduced with respect to the yield levels of other $p$-nuclides.
The ratios between $p$-nuclei and iron in the ejecta are larger than
the solar ratios by a factor of 1.2.  We also compare the yields of
oxygen, iron, and $p$-nuclides in SNe~Ia and SNe~II, and suggest that
SNe~Ia could make a larger contribution than SNe~II to the solar
system content of $p$-nuclei.
\end{abstract}

\keywords{Galaxy: abundances --– nuclear reactions, nucleosynthesis,
abundances –-- supernovae: general}

\section{INTRODUCTION}\label{sec1}

Stable nuclides of atomic number $Z \geq 34$ located
at the neutron deficient side of the $\beta$-stability line are
classified as $p$-nuclei~\citep[e.g.][]{lam92,mey94,arn03}.  They consist
of 35 nuclides. The production process of these $p$-nuclei
is commonly called the $p$-process.  The process through the 
photodisintegration reactions on the pre-existing heavy nuclides
is, especially, referred to as the $\gamma$-process~\citep{woo78,how91}.  These
nuclides are observed only in the solar system, since the abundances are
very small, typically 1\% or less in the isotopes of element (by number). 
Some primitive meteorites which were not in equilibrium with the bulk of
the solar system materials are also found to contain very poor
$p$-nuclei~\citep{and89}.

Although nucleosynthetic processes to produce $p$-nuclei have been studied
for many astrophysical sites, the core-collapse supernova (SN) of massive
stars with an H-rich envelope, i.e., a Type II supernova (SN~II), has
been considered as a plausible
site~\citep[e.g.,][]{woo78,ray90,pra90,ray95,cos00,rau02,iwa05,hay04,hay06,hay08}.
During the SN explosion, the $\gamma$-process plays an important role 
in producing $p$-nuclei in the O- and Ne-rich
layers.

Many studies of SNe~II have shown that about a half of $p$-nuclides are
reproduced in the proportion of the solar $p$-abundance.  However, 
there remain two unresolved problems.  First, $^{92,94}$Mo, $^{96,98}$Ru, $^{115}$Sn, and 
$^{138}$La are largely underproduced compared with the distribution 
of the solar $p$-abundances.
Second, the contribution of $p$-nuclei from SNe~II to 
their galactic evolution has been found to be smaller than that of $^{16}$O, the
main product of SNe~II~\citep{pra90,ray95}. These results led
them to conclude that SNe~II could not be responsible for the whole 
content of solar $p$-nuclei.

The $p$-process nucleosynthesis in a supercritical accretion disk around
a compact object~\citep{fuj03} and in a jet-like explosion~\citep{nis06} has
been studied.  
\citet{hof96} suggested that nucleosynthesis in the neutrino-driven
wind following the delayed explosion can produce light $p$-nuclei, i.e.,
$^{74}$Se, $^{78}$Kr, $^{84}$Sr, and $^{92}$Mo.
Recently, neutrinos emitted from the collapsed core has been found to
largely contribute to the production of $^{92,94}$Mo, $^{96,98}$Ru, and
other light $p$-nuclei~\citep{pru05,fro06a,fro06b,pru06,wan06}.  This
contribution might resolve the 
underproduction of $^{92,94}$Mo and $^{96,98}$Ru in the O- and Ne-rich
layers in SNe~II.

\citet{cos00} have shown 
that the second problem of underproduction of $p$-nuclei could be
solved if the $^{22}$Ne($\alpha,n$)$^{25}$Mg reaction rate
was larger by a factor of $\sim 10$--$50$ in the
temperature range of the $s$-process during core He burning.  However,
the uncertainty they expected is too large and now
questioned~\citep{jae01,koe02,kar06}.  On the other hand, the second problem might be
reconciled by considering the contribution of more energetic SNe
(hypernovae) to the production of $p$-nuclei~\citep{iwa05,hay08}.  

It is less likely that uncertainties of the nuclear reaction rates
such as ($\gamma$,$n$), ($\gamma$,$\alpha$), and
($\gamma$,$p$) photodisintegrations (and their inverse reactions) for nuclei
heavier than iron give a significant impact on a $p$-process yield as seen from the 
sensitivity of the $p$-nuclei yields on reaction rates ~\citep{rap06}.  \citet{dil08} have performed $p$-process calculations for the same SN II model
as in~\citet{ray95} by utilizing most recent stellar ($n$,$\gamma$) cross
sections.  They found that 
overproduction factors for almost all $p$-nuclides decreased by, on
average, 7\% and that the largest deviation is the reduction in $^{156}$Dy by
39.2\% in comparison with the result in~\citet{rap06}.

The $p$-process has also been suggested to occur in the
outermost layer of the exploding carbon-oxygen (CO) white dwarf (WD)
which is presumed to be a Type~Ia supernova~\citep[SN Ia; ][]{how91,how92,gor02,kus05,gor05,arn06}.  \citet{how91}
calculated the $p$-process in a simple parametric model. 
In this model, it is assumed that the $s$-process produces seed nuclei
with the mass number higher than 90 prior to the explosion.  The
high density environment during the passage of the shock wave
makes proton capture reactions efficient in producing light
$p$-nuclides, while the $\gamma$-process makes heavier $p$-nuclides.
They reproduced the solar-like distribution of $p$-nuclei.
\citet{how92} calculated the $p$-process nucleosynthesis in the 
delayed-detonation (DD) model and obtained an abundance pattern where the
abundances of lighter $p$-nuclei are relatively large.  When the solar
abundance is used for the initial composition, the result implies that
SN~Ia has very small contribution to the galactic content of $p$-nuclei.
However, when the $s$-process is assumed to occur as in the
solar metallicity asymptotic giant branch (AGB)
star, a sufficient yield of $p$-nuclei is
obtained.  In this case, the underproductions of Mo
and Ru are reduced, compared to those in SNe~II.  Those
authors concluded that contribution of SN Ia on Galactic $p$-nuclei was
still uncertain. It is, therefore, worth further studies.

\citet{gor02} studied the $p$-process nucleosynthesis in the
one-dimensional He-detonation model for a sub-Chandrasekhar mass CO WD, where the
$p$-process would proceed in the accreting He layer.
For the initial seed with solar abundances in accreting He-rich materials,
the overproduction of
Ca-to-Fe nuclei has been found to be a factor of $\sim 100$ with respect to
$p$-nuclei. This result means
that the He detonating sub-Chandrasekhar mass CO WD model is not an
efficient site for the synthesis of 
$p$-nuclei.  Meanwhile, they have shown that by increasing the abundances of
the
initial heavy seeds with the solar composition by a factor of 100
the $p$-nuclei yields are comparable to those of Ca-to-Fe nuclides.  The
resulting yields of $p$-nuclei in their one-dimensional model have been confirmed
to be very similar to those calculated by using a three-dimensional explosion
model~\citep{gor05}.

In light of the theoretical underproduction of $^{92,94}$Mo and
$^{96,98}$Ru, a precise measurement of terrestrial abundances of those
isotopes is needed.  Recently,~\cite{lae08} has shown that the abundances
of $p$-nuclei of Mo and Ru were, on average, 4.3\% lower than presently
known ones.  Unfortunately, the large underproduction problem of
the Mo and Ru isotopes still remains.

In this paper, we have adopted a carbon-deflagration model (W7) for 
SN~Ia~\citep{nom84} and analyzed the $p$-process
nucleosynthesis. The used trajectories of temperature and density in
exploding WD are largely different from those in the DD
model by \cite{how92} and in the parametric model by \cite{how91}. It
is, thus, important to investigate the C-deflagration model as one of
the possible sites for the $p$-process.
The structure of this paper is as follows.
The adopted SN model, nuclear reaction network,
and initial compositions are described in Section~\ref{sec2}.  The results
for some cases of initial compositions are analyzed in Section~\ref{sec3}.
Finally, we present conclusions in Section~\ref{sec4}.

\section{INPUT PHYSICS}\label{sec2}
\subsection{The Supernova Model}

The adopted SN model is W7 in Nomoto et al. (1984).  
The accreting WD with the initial mass of $1.0~M_{\sun}$ has been cooled
down for $5.8\times 10^8$~yr before the onset of mass
accretion. This WD has the composition of  
$X(^{12}{\rm C})=0.475$, $X(^{16}{\rm O})=0.5$, and 
$X(^{22}{\rm Ne})=0.025$. The WD mass increases at the accretion rate of 
$\dot{M}=4\times 10^{-8}\ M_{\sun}~{\rm yr}^{-1}$. 
When the WD mass approaches $M_{\rm WD}=1.378~M_{\sun}$, carbon burning
is ignited at the center. This forms a C-deflagration
wave which propagates outward. The released nuclear energy of 
about $10^{51}$~erg exceeds the binding 
energy of the WD so that the whole star is exploded. 
We use the time variations (trajectories) of temperature and density in
the exploding layers~\citep{nom84} in order to
calculate the production of $p$-nuclei.

\begin{figure}
\begin{center}
\includegraphics[angle=-90,scale=0.3]{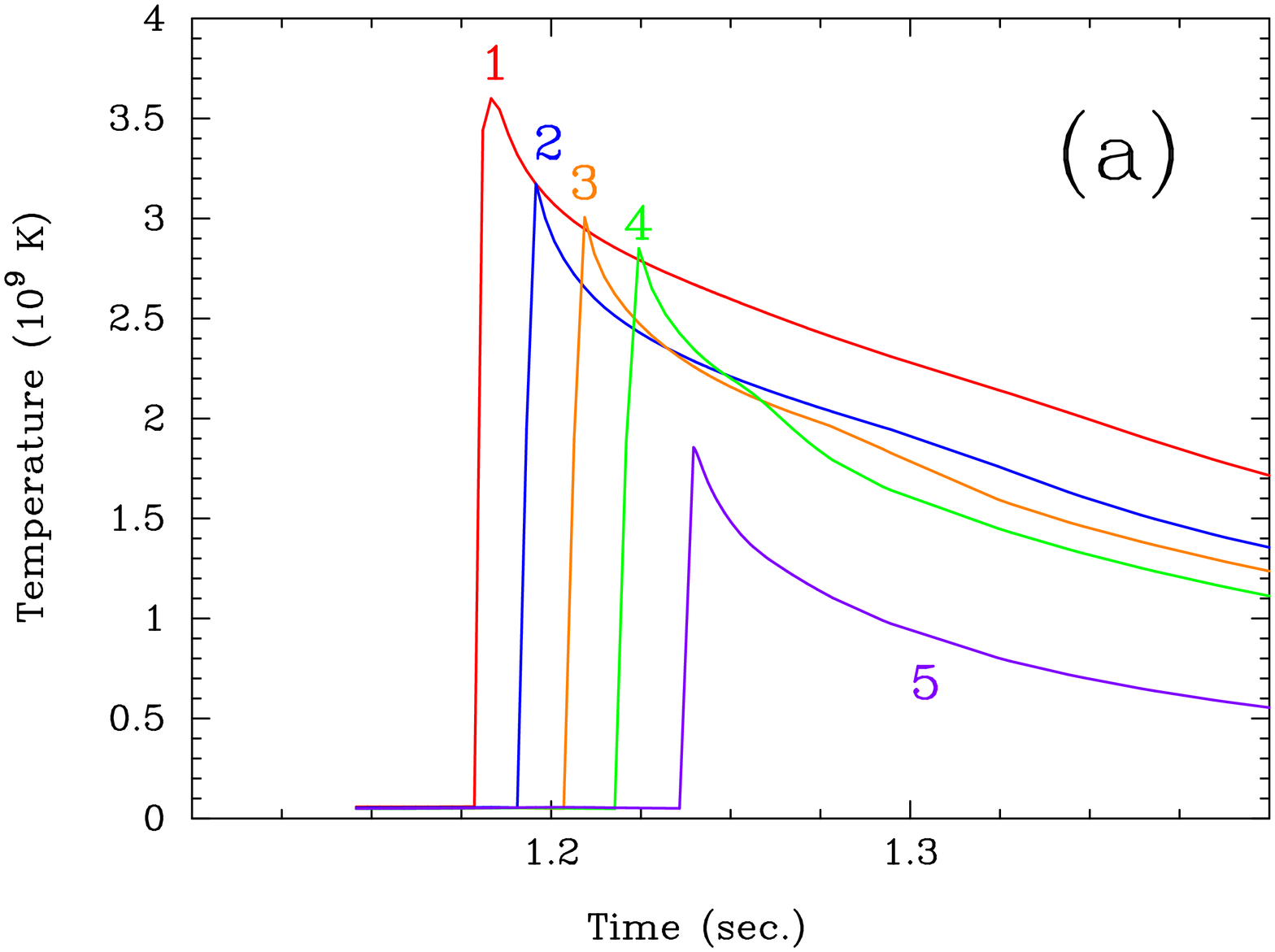}
\includegraphics[angle=-90,scale=0.3]{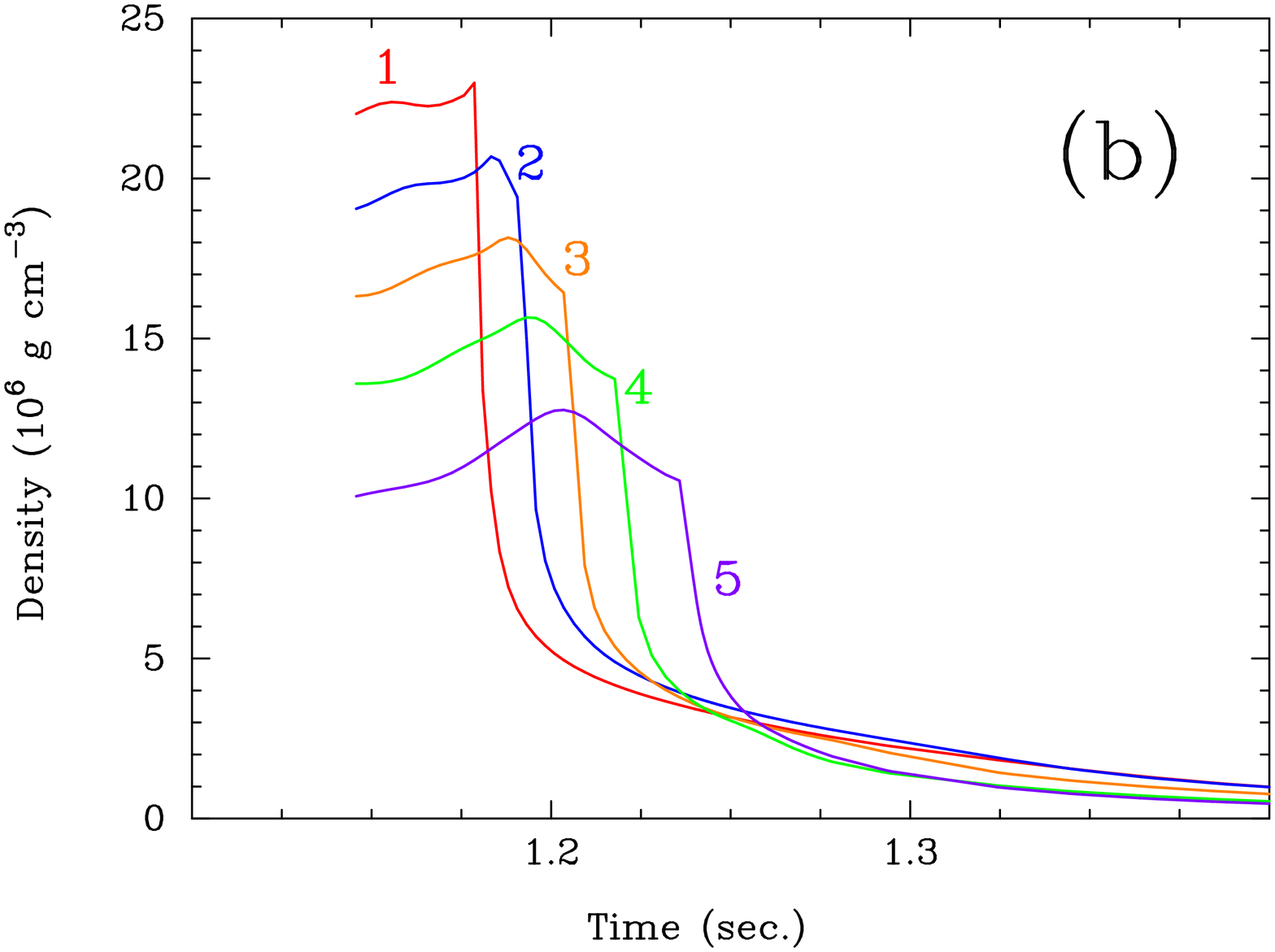}
\end{center}
\caption{Trajectories of temperature (a) and density (b) as a
 function of time in the W7 model. The labels refer to the trajectories in
 the layers with
 the peak temperatures in units of $10^9$~K, i.e., $T_{{\rm m},9}=T_{\rm m}/(10^9~{\rm K})=3.60$ (1), 3.17 (2),
 3.01 (3), 2.85 (4), and 1.86 (5).\label{fig1}}
\end{figure}

We calculate the $p$-process nucleosynthesis 
in the heated layers where a peak temperature $T_{\rm m}$ has the range of   
$T_{{\rm m},9}= 1.86$--$3.60$ (in units of $10^9$~K). The corresponding
peak densities are 
$\rho_{\rm m}=1.28\times 10^7$ to $2.24\times 10^7\ {\rm g~cm^{-3}}$.
These layers are located at mass coordinates of
$1.143<M_r/M_{\sun}<1.280$ and undergo
explosive carbon and neon burning at a passage of the deflagration wave.
The representative temperature and density trajectories are shown in
Figures~\ref{fig1}(a) and \ref{fig1}(b), respectively.  
It should be noted that the yields of $p$-nuclei are very
sensitive to the temperature and density trajectories of exploding WD.  
The DD and C-deflagration (W7) models show completely
different characteristics of trajectories. In the W7 model, the
decreasing timescale of temperature in 
regions of $T \gtrsim 2\times 10^9$~K relevant to the $p$-process is
$0.15$--$0.3$~s, depending on mass shells.  This timescale is 
$\sim 1.5$--$2$ times longer than in the DD model by \cite{how92} and 
$2$--$4$ times shorter than in the parametric model with the assumption
of $e$-folding time (0.6~s) by \cite{how91}.  
During the $p$-process nucleosynthesis, the density in W7 
model is higher than in the DD model and in the assumption in \citet{how91}.
The difference in density is also important and affects the yields of
light $p$-nuclei resulting from proton capture reactions.  

\placefigure{fig1}

\subsection{Nuclear Reaction Network and Initial Composition}
The $p$-process nucleosynthesis is calculated by using the nuclear
reaction network, in which 2565 nuclei from neutron and proton to
Polonium ($Z=84$) are combined with neutron,
proton, $\alpha$-induced reactions, and their inverses. We use the nuclear
reaction rates based on the experiments and the Hauser-Feshbach
statistical model, 
NON-SMOKER~\citep{rt00}. The theoretical and experimental $\beta$-decay
rates are adopted from the REACLIB database~(F.-K. Thielemann 1995,
private communication), which are
supplemented by the theoretical rates of \cite{mnk97}. 
We take the solar abundances from~\cite{and89} to calculate the
overproduction factor, $X/X_\sun$, of the produced $p$-nuclei, where $X$
is the mass fraction.

The seed abundances are important to consider the nucleosynthesis by the
$p$-process in SNe~Ia. In the W7 model, materials
with solar compositions are accreted onto the surface of the CO WD.  
The accreted H-rich materials are transformed into He
through the CNO cycle of H-burning. The main product of the CNO
cycle, $^{14}$N, is left in the He-rich region and is processed to
$^{22}$Ne through the $\alpha$ capture reactions. The main nuclei in the
CO WD result in $^{12}$C, $^{16}$O, and $^{22}$Ne.  

The accretion rate for the W7 model is high enough to avoid the
occurrence of an He detonation in the sub-Chandrasekhar mass
stage~\citep{nom82a,nom82b}.  Instead, He burning is ignited in a very
thin shell 
at the base of the He-rich layer where electrons are partially
degenerate. The He shell burning becomes thermally unstable in
these conditions, which are 
similar to those in the He-burning shell of an AGB
star. This results in the repeated occurrence of thermal runaway of He
burning (He shell flash).

For enhanced $p$-process to occur in W7 model, $s$-process must occur when
the WD mass is 1.143--1.280 $M_\sun$.  For such high WD mass (as well as
the high-mass C+O core of AGB stars), the temperature of the He-burning
layer is high enough for the $^{22}$Ne($\alpha,n$)$^{25}$Mg reaction
to take place \citep[e.g.][]{fuj77,tru77,str00}.  
There is an important difference between the accreting WD and the AGB
star, i.e., the mass of the H-rich layer in the accreting WD is much
smaller than the AGB star and thus the entropy of the H-layer is much
smaller in the WD.  Therefore, the mixing of H-rich material into the
He layer is more easily to occur \citep{sug78}.
Therefore, it is naturally expected that the
$s$-process nucleosynthesis proceeds through the neutron source
reactions of $^{22}$Ne($\alpha,n$)$^{25}$Mg as well as of 
$^{13}$C($\alpha,n$)$^{16}$O and finally creates a large 
amount of heavy nuclei. 

\begin{figure}
\begin{center}
\includegraphics[angle=-90,scale=0.3]{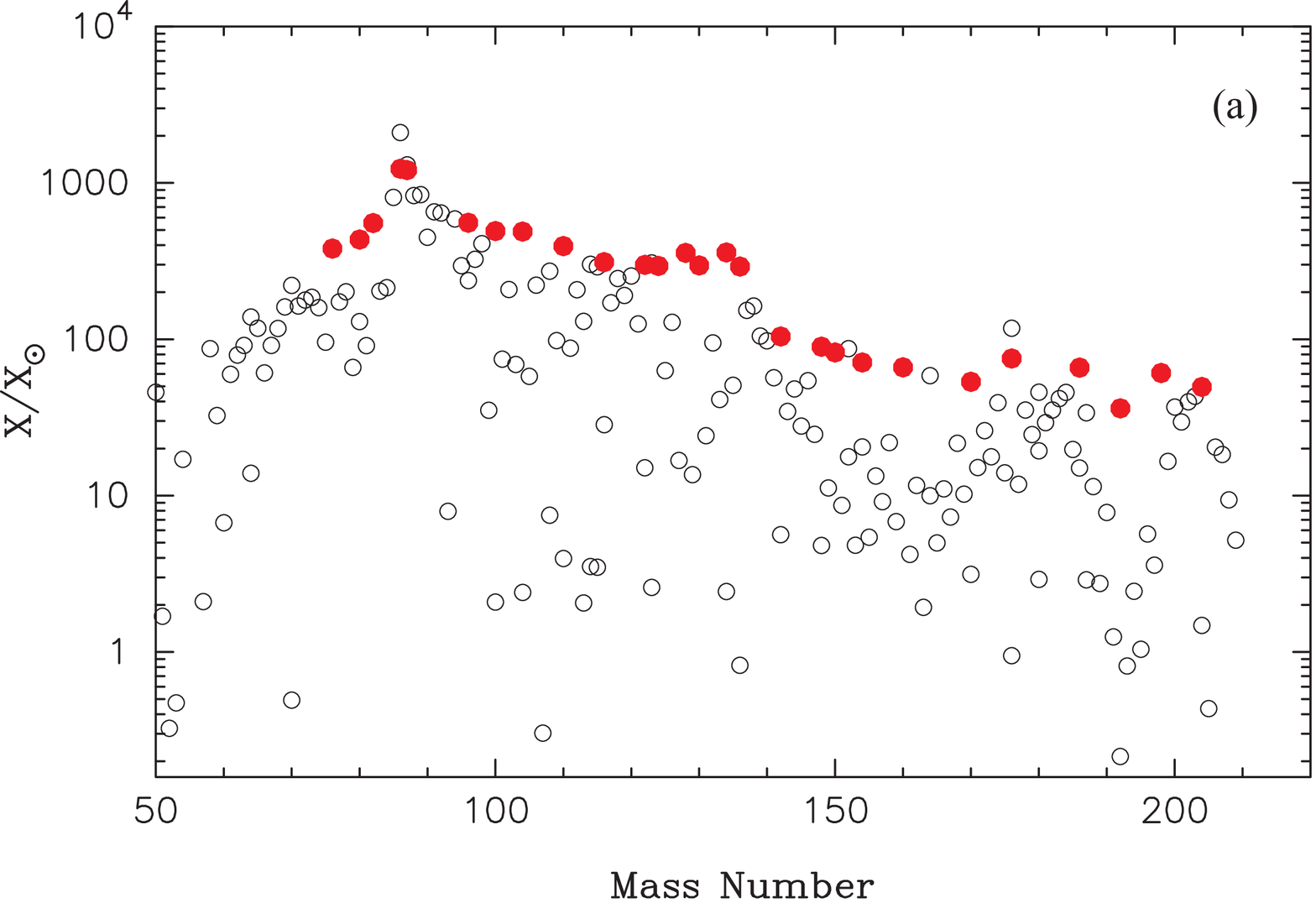}
\includegraphics[angle=-90,scale=0.3]{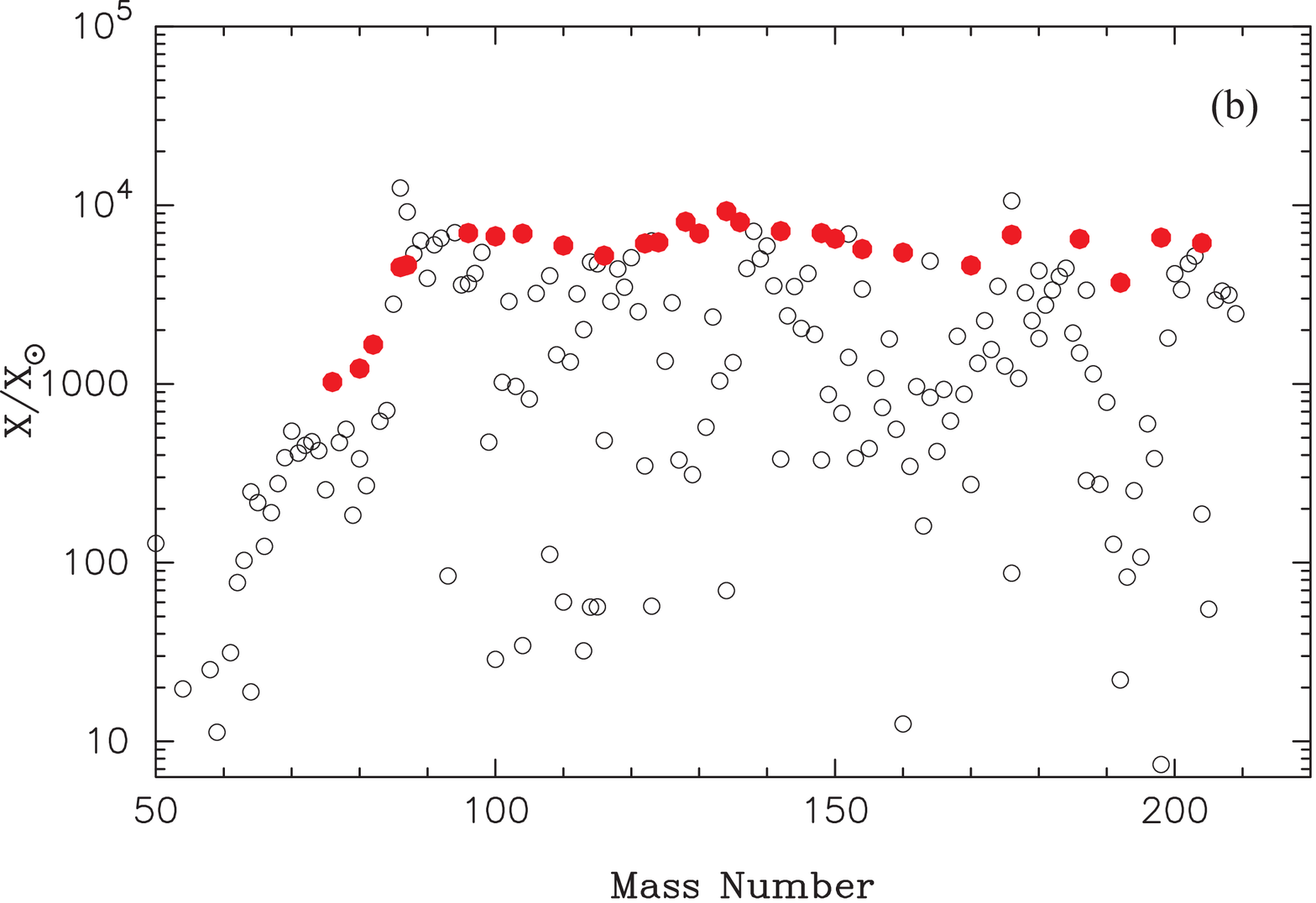}
\end{center}
\caption{Distributions of initial seed abundances relative to solar for
 Case~A (a) and B (b) (Table \ref{tab1}), respectively.  Filled circles indicate $s$-only
 nuclides.\label{fig2}}
\end{figure}

The efficiency of the $s$-process in the
accreting CO WD remains uncertain, and thus, we calculate the
initial seed 
abundances for the $p$-process by using the canonical $s$-process
model~\citep{how86,aok03,ter06}. The nuclear reaction network for the
$s$-process includes 602 nuclei connected with the neutron capture
reactions, whose rates are taken from \cite{bao00} and \cite{rt00}, and the
$\beta$-decays, whose rates are adopted from \cite{tak87} and REACLIB.  Nuclei above $^{32}$S are included in the seed composition
since the $s$-process nucleosynthesis is dominant. 
In this investigation, we assume two sets of initial
seed abundances with mean neutron exposure $\tau_0=0.15$ (Case~A) and
0.33~mb$^{-1}$ (Case B) for
the $p$-process. Figure~\ref{fig2} shows the seed abundance distributions
normalized to the solar abundances, and Tables~\ref{tab6} and
\ref{tab7} show the seed abundances (by mass) in A and B,
respectively. The seed abundance of B is 
fitted to the solar $s$-only nuclei,  
and thus, the distribution is similar to the solar $s$-distribution.  
The abundances of B are higher than those of A because the
larger production efficiency is needed to get the constant 
$s$-distribution normalized to solar.

\placefigure{fig2}

The He-exhausted core is composed of mainly $^{12}$C and $^{16}$O. The C/O
abundance ratio (by mass) is assumed to be 0.95 (Case~A1 in
Table~\ref{tab1}) in the W7 model. However, the C/O 
ratio should be changed by, e.g., the adopted reaction rate of
$^{12}$C($\alpha,\gamma$)$^{16}$O which has a large uncertainty in the 
current experiments at a low energy~\citep[e.g.,][]{mak07}. 
We investigate the influence of the different C/O ratio on the
$p$-process nucleosynthesis by changing the C/O ratio from 0.56
(Case~A2) to 2.55 (Case~A3).
The $^{22}$Ne abundance left after the
$s$-processing is also uncertain. The high abundance of $^{22}$Ne
may affect the $p$-process flows by the production of neutrons through
$^{22}$Ne($\alpha,n$)$^{25}$Mg reaction. In order to investigate the
influence, we calculate the case in which no $^{22}$Ne abundance is left
after the He shell flashes as an extreme case (Case~A4). The $^{12}$C,
$^{16}$O, and $^{22}$Ne  
abundances for the adopted cases are summarized in Table~\ref{tab1}.

\placetable{tab1}

\section{RESULTS}\label{sec3}

\subsection{Case A1}

Here, we show the result of Case~A1, which is considered as the
standard case in this investigation. 

\subsubsection{Abundance Variations of Light Particles, $p$, $n$,
   $\alpha$}

\begin{figure}
\begin{center}
\includegraphics[angle=-90,scale=0.3]{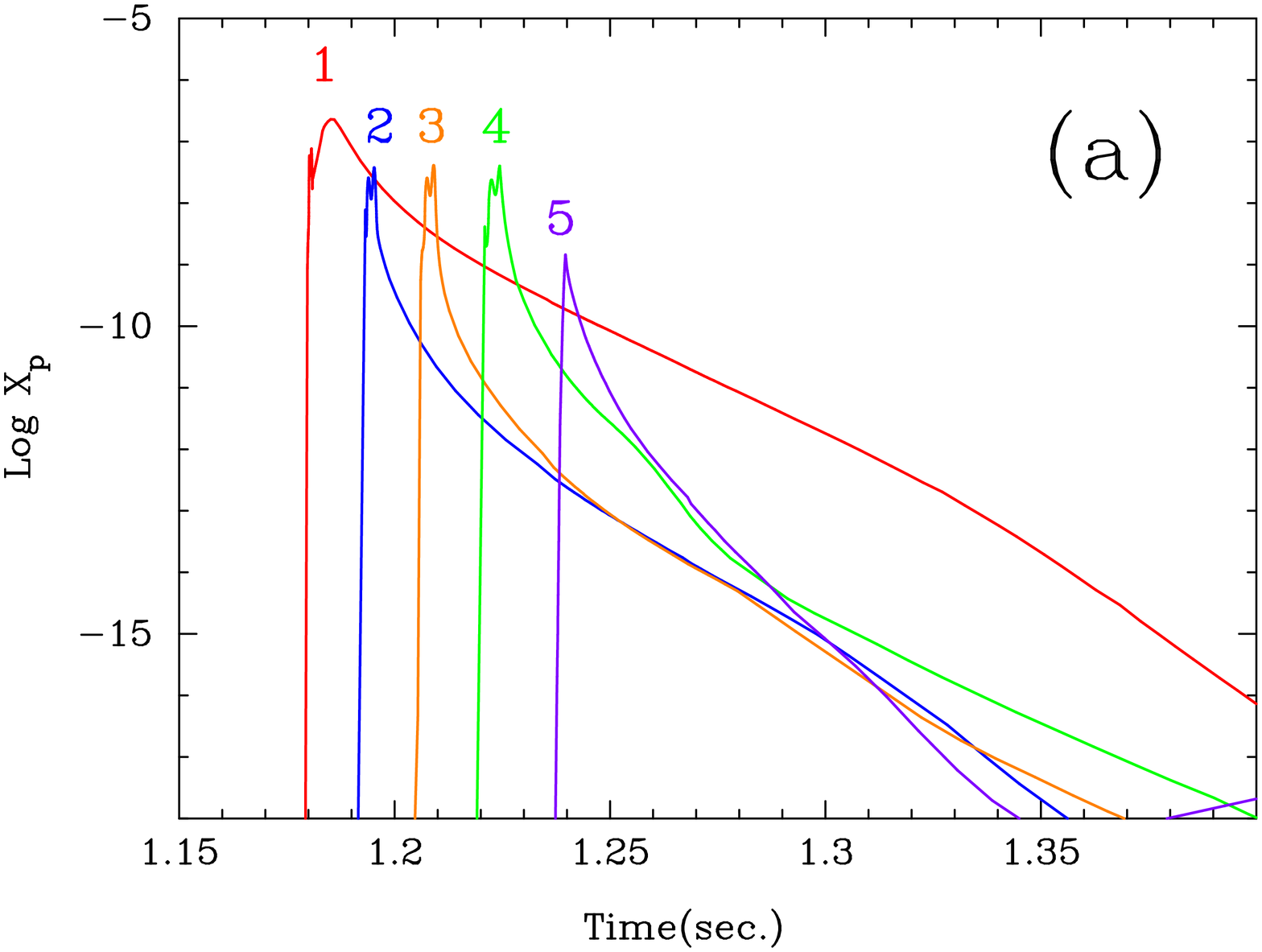}
\includegraphics[angle=-90,scale=0.3]{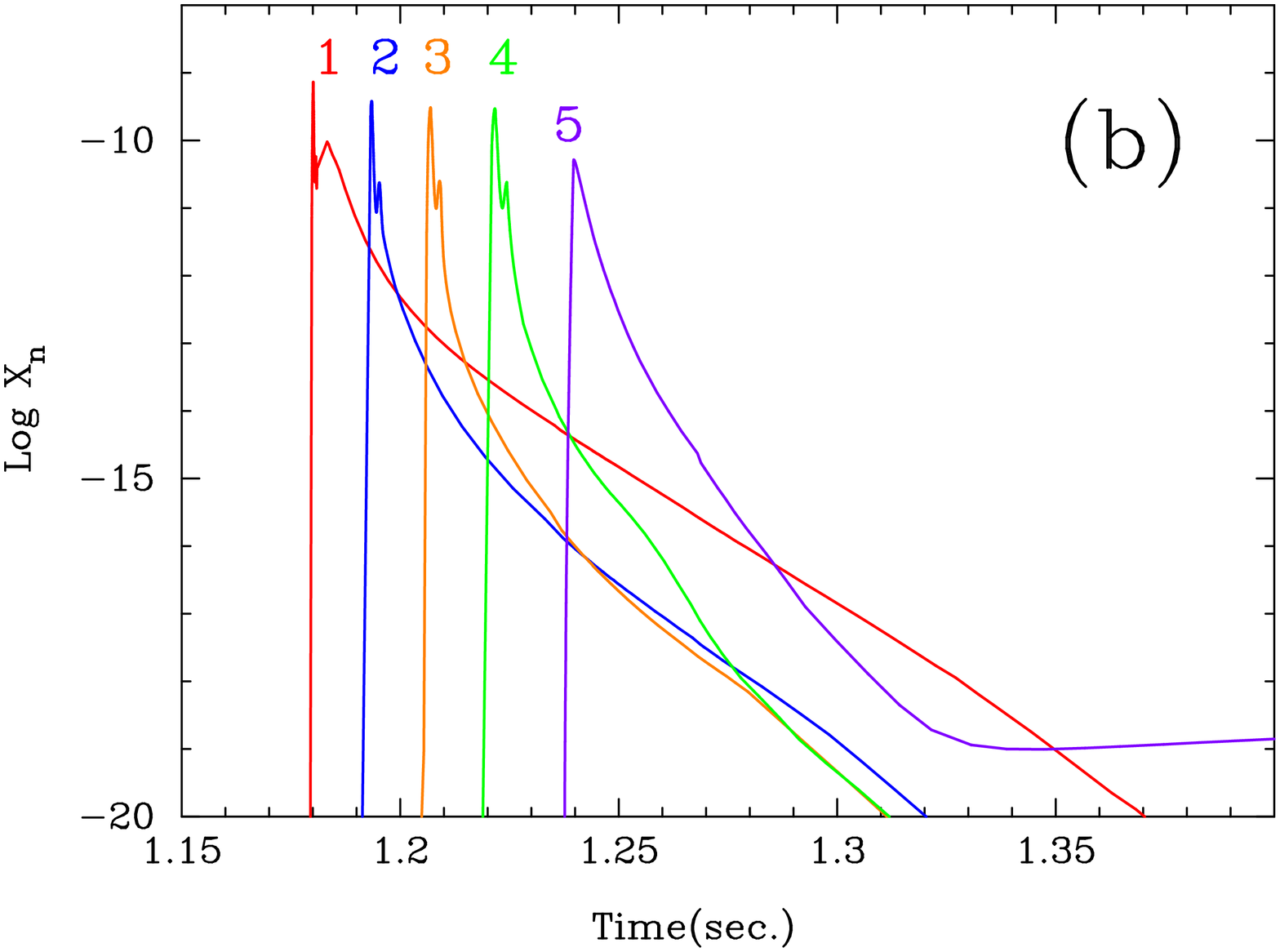}
\end{center}
\begin{center}
\includegraphics[angle=-90,scale=0.3]{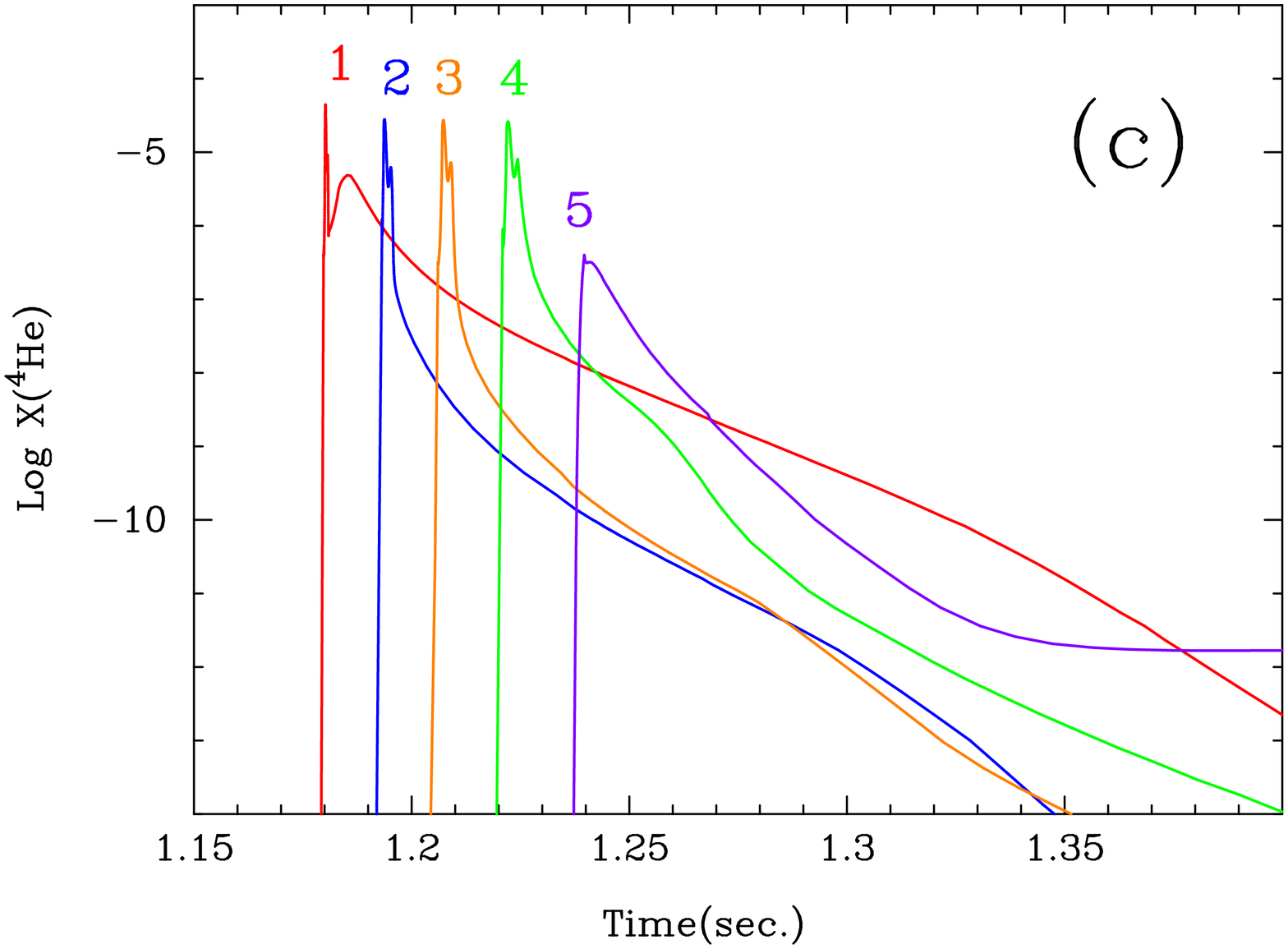}
\includegraphics[angle=-90,scale=0.3]{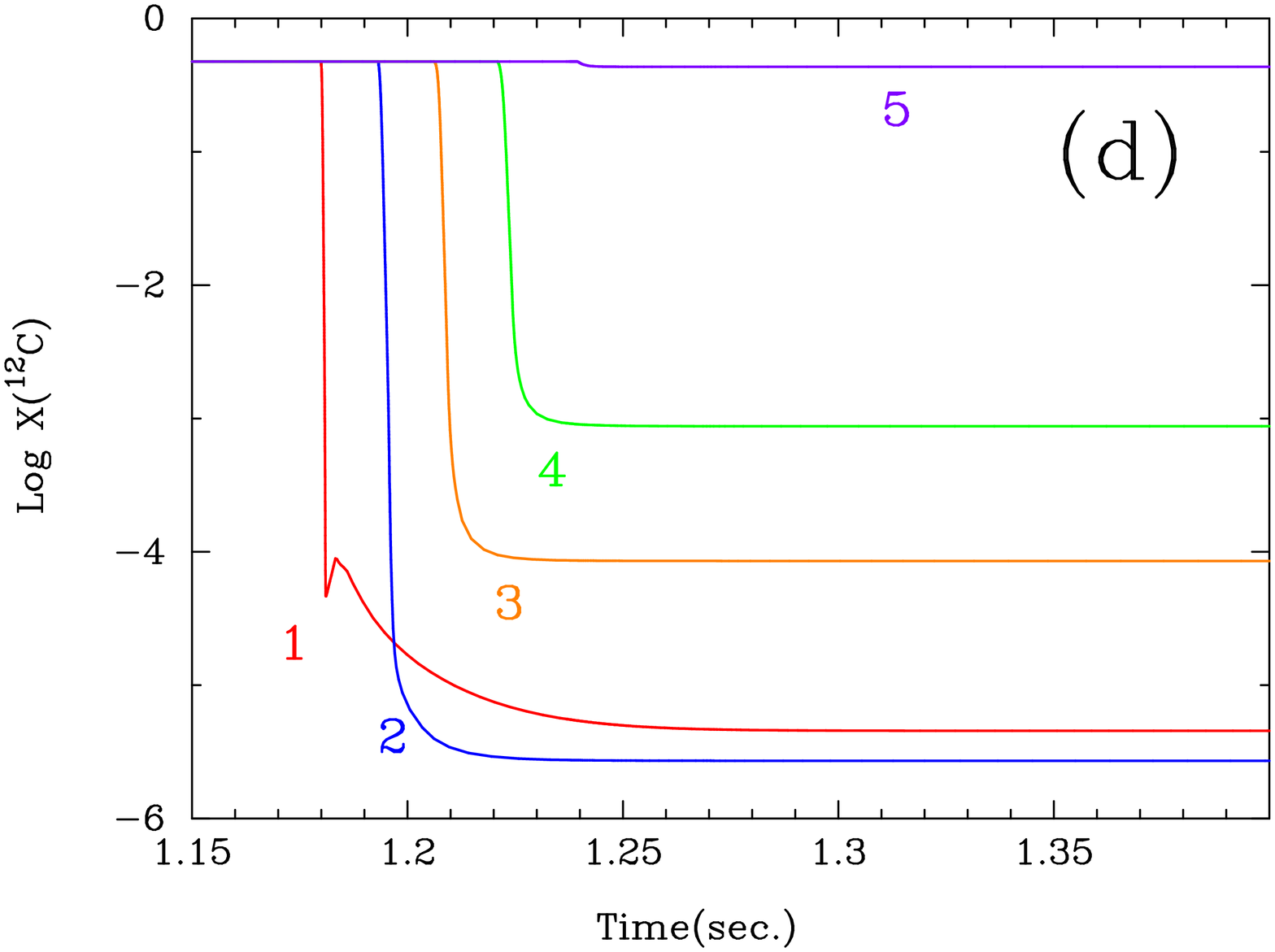}
\end{center}
\begin{center}
\includegraphics[angle=-90,scale=0.3]{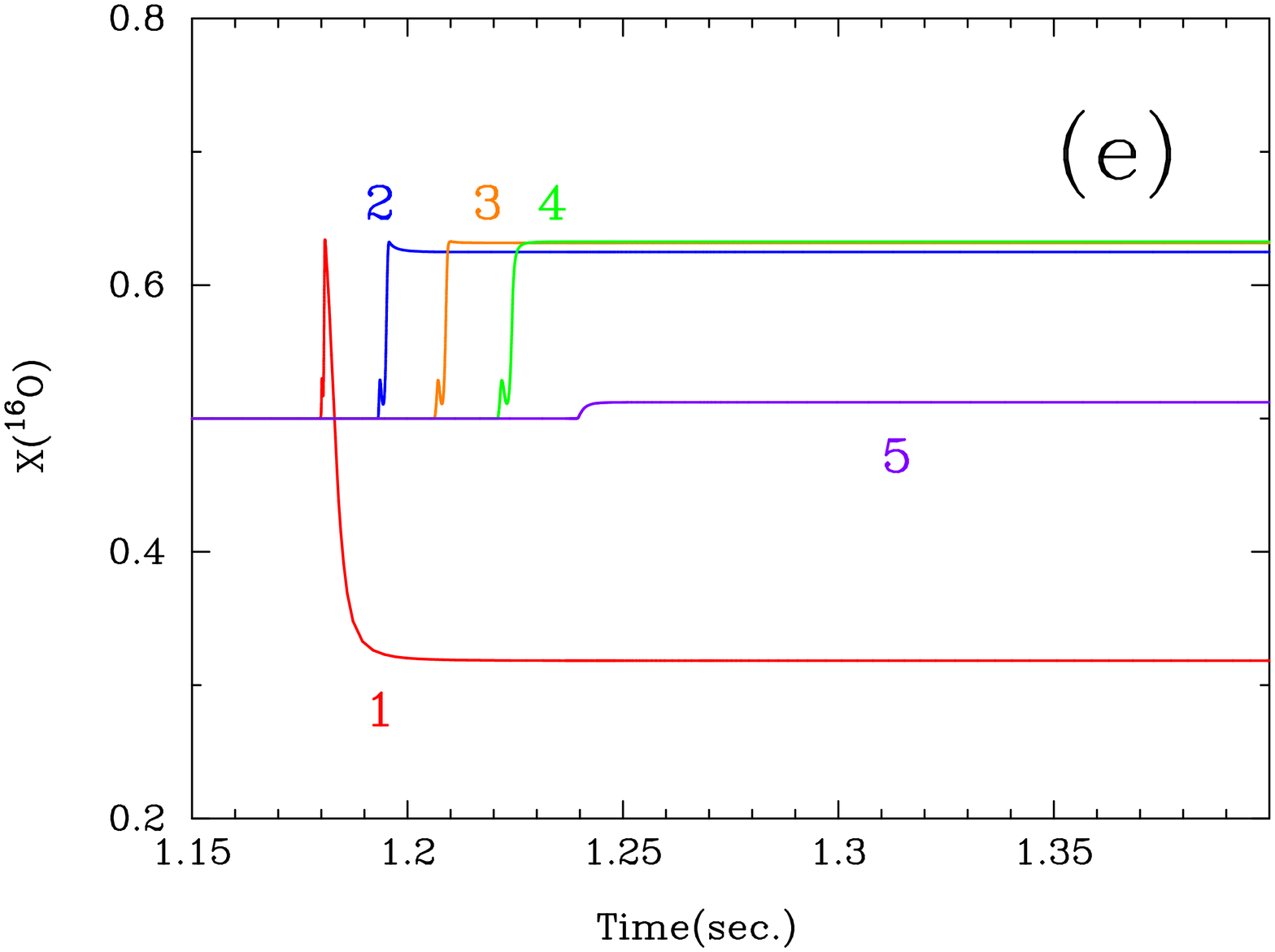}
\includegraphics[angle=-90,scale=0.3]{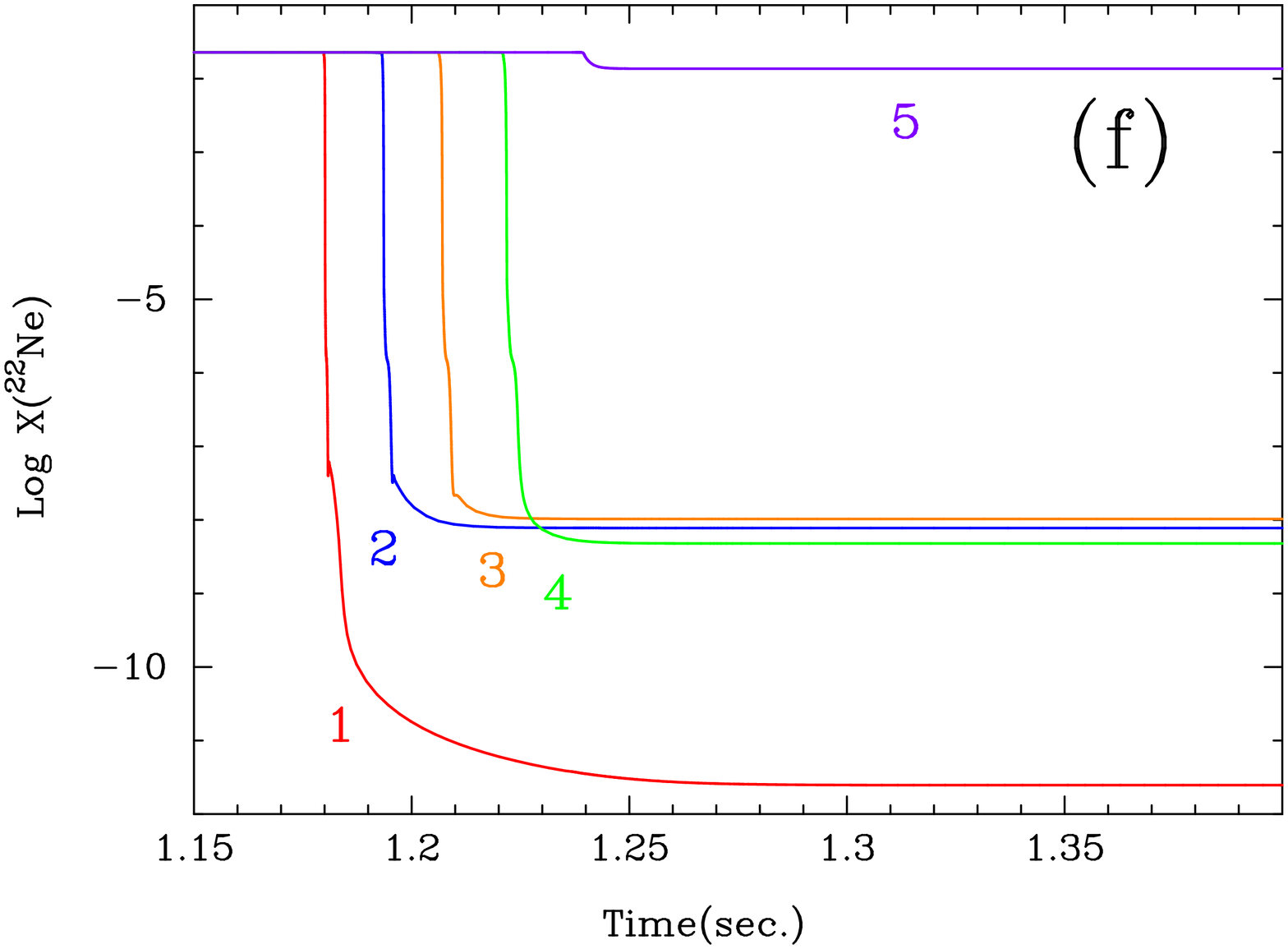}
\end{center}
\caption{Time evolution of mass fractions of proton (a), neutron (b),
 $^{4}$He (c), $^{12}$C (d), $^{16}$O (e), and $^{22}$Ne (f) during the
 explosion in the five layers shown in Figure\ \ref{fig1}.\label{fig3}}
\end{figure}

Figure~\ref{fig3} represents the abundance variations of 
proton, neutron, $^4$He, $^{12}$C, $^{16}$O, and $^{22}$Ne in each
trajectory as a function of time after the onset of explosion.  
In layer 1 where the peak temperature attains to 
$T_{{\rm m},9}=3.6$, $^{16}$O+$^{16}$O fusion reaction 
starts to destroy $^{16}$O as the temperature increases rapidly.  
Layers 2--4 are the most interesting region as the main site of
$p$-process, where explosive C and Ne burning successively occurs.  
$^{20}$Ne first produced is photodisintegrated, and
thus, $^{16}$O is left (Figure~\ref{fig3}(e)).
The explosive C-burning partially operates in layer 5, but the layer
is not important for the $p$-process. $^{22}$Ne is burnt by
($\alpha,\gamma$) and ($\alpha,n$) reactions in the layers 1--4. 
The peak abundances of
$\alpha$-particle are $X_{\alpha}\sim 10^{-5}$ in all layers, which are
larger than those in the $p$-process in SNe~II~\citep[e.g.,][]{pra90}.
In the SN II model of \cite{pra90},
$X_{\alpha}\sim 10^{-5}$ is realized by the efficient
photodisintegration of $^{20}$Ne at $T_9\sim 3$, but it decreases to
$X_{\alpha}\sim 10^{-7}$ as the peak temperatures decrease. 
In the SNe~Ia, $\alpha$ particle is also produced by the
$^{12}$C($^{12}$C, $\alpha$)$^{20}$Ne($\gamma$, $\alpha$) in spite of no
initial $^{20}$Ne abundance. 
The peak proton and neutron abundances are $X_p\sim 10^{-7}$
and  $X_n\sim 10^{-10}$, respectively, which are also larger than those
in SNe~II.  

\placefigure{fig3}

In the present model, neutron is produced mainly by the
$^{22}$Ne($\alpha,n$) reaction. The peak neutron number 
density is $N_n\sim 10^{22}$--$10^{23}$ cm$^{-3}$, which is almost equivalent to
that in the He-detonation of sub-Chandrasekhar-mass model~\citep{gor02}.
On the other hand, the peak proton mass fraction 
is about five orders of magnitude smaller than that in the He-detonation
model ($X_p\sim 6\times10^{-3}$ in the He-rich layer with 
$T_{\rm m,9} \gtrsim 3$). 
In the He-detonation model, the plenty of $\alpha$-particles are present
in the region where the $p$-process occurs 
and lead to the production of $\alpha$-elements such as $^{40}$Ca and
$^{44}$Ti.  Further radiative $\alpha$-captures bring the nuclear flow to the
proton-rich side of the $\beta$-stability line, and then are followed by
($\alpha$,$p$) reactions. This results in the very high proton mass
fraction.
In the C-deflagration model, on the contrary, the overproduction of
proton does not take place, since the $p$-process occurs in outer
CO-rich layers as explained below.

\subsubsection{Nuclear Flow}

The general production trends of various $p$-nuclei are understood
through the analysis of nuclear flows.  First, we see a nuclear flow in
the layers with a high peak temperature.  After a passage of the
deflagration wave the high neutron density produced
brings almost all the seed nuclei to the neutron-rich side through
($n,\gamma$) reactions.  In this C-deflagration
model, the abundances move to more
neutron-rich region than in 
SNe~II.\footnote{We checked the $p$-process calculation in the following
approximate model of SNe~II.  The temperature and density are given by $T(t)=T_{\rm
m}\exp(-t/3\tau_{\rm ex})$ and $\rho(t)=\rho_{\rm m}\exp(-t/\tau_{\rm
ex})$, respectively, where $\tau_{\rm ex}$ is the expansion timescale
taken to be 0.446~s and 1~s, and $\rho_{\rm m}=10^6$~g~cm$^{-3}$.  This
setup is the same as in~\citet{ray90}.}

(1) When the temperature
exceeds $T_9\sim2$, ($\gamma,n$) reactions bring back the
nuclear flow to the $\beta$-stability line and further to the
neutron-deficient region.  There is a little flow
which results in a leakage to lower $Z$ through ($\gamma,p$) and
($\gamma,\alpha$) photodisintegrations.

(2) Furthermore, when the
temperature exceeds $T_9\sim 3$, ($\gamma,p$) and ($\gamma,\alpha$)
reactions become dominant, which drives nuclear materials from the
neutron-deficient, high $A$ region down toward the iron-peak element region.
This flow to the iron-peak elements is terminated, since the
photodisintegrations freeze out as the temperature of the heated layers
decreases. 

\begin{figure}
\begin{center}
\includegraphics[angle=-90,scale=0.5]{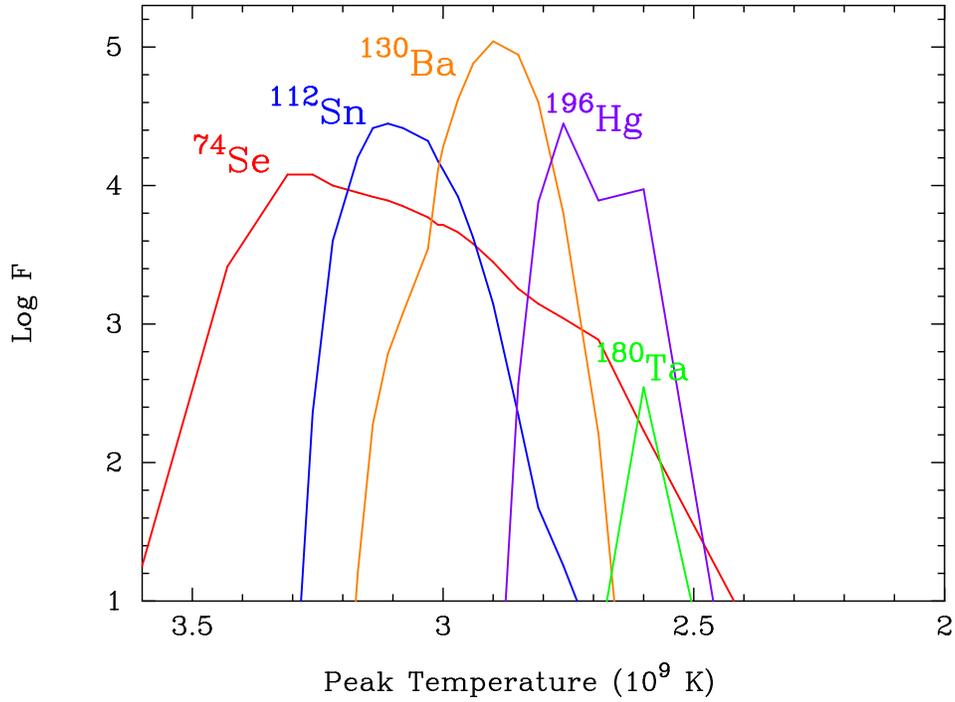}
\end{center}
\caption{Dependence of the production of representative $p$-nuclei
 ($^{74}$Se, $^{112}$Sn, $^{130}$Ba, $^{180}$Ta, and $^{196}$Hg) on
 the peak temperature $T_{\rm m}$.\label{fig4}}
\end{figure}

(3) In the layer with $T_{{\rm m},9}\simeq 2.6$, the
($\gamma,n$) reactions are predominant to the seed nuclei with neutron
number $N>82$ and create the production peak for most of the $p$-nuclei
with $N>82$. This result is seen for 
$^{196}$Hg in Figure~\ref{fig4}. $^{180}$Ta, which is one of the rare
nuclei in nature, is produced directly
through the photodisintegration of $^{181}$Ta in Figure~\ref{fig4}.  We
note that the production of $^{180}$Ta might become efficient even in
layers with lower peak temperatures~\citep[$T_{{\rm m},9}= 1.8$--$2.6$; ][]{pra90}. Unfortunately, we 
cannot investigate the $p$-process nucleosynthesis in such
layers.\footnote{A mesh zoning in the outer layer of the W7 model was
sparse. This is because this region is not so important for main
nucleosynthesis in SN~Ia.}

(4) In a somewhat higher peak temperature region 
($T_{{\rm m},9}\sim 2.7$--$2.9$),
($\gamma,n$) reactions drive the nuclear flow to a sufficiently
neutron-deficient region, and then, ($\gamma,p$) and ($\gamma,\alpha$)
reactions get effective to destroy seed nuclei with $N>82$. Large
amounts of heavy $p$-nuclei with $N>82$ are produced in this
peak temperature range from $\beta^+$-decays of unstable isobars 
after freezing-out of the nuclear reactions. For example, the second
production peak is found for $^{196}$Hg in Figure~\ref{fig4}. 
The production of intermediate-mass $p$-nuclei with $50<N<82$
(especially, $^{113}$In, $^{115}$Sn, $^{120}$Te, $^{126}$Xe, $^{132}$Ba,
$^{138}$La, and $^{136,138}$Ce) already proceeds by ($\gamma,n$)
photodisintegrations in this region. 

(5) In the layers where the peak temperatures reach 
$T_{{\rm m},9}\sim 2.9$--$3.2$, most of the intermediate-mass seed nuclei
experience photodisintegrations. Then, the intermediate-mass $p$-nuclei
are abundantly produced mainly through the $\beta^+$-decays of
unstable neutron-deficient isobars.

(6) In the layers with $T_{{\rm m},9}= 3.2$--$3.3$ 
$^{92,94}$Mo, $^{96,98}$Ru and $^{102}$Pd show production peaks, but
the abundances rapidly decrease when the peak temperature exceeds 
$T_{{\rm m},9}= 3.3$. Finally, all the $N>50$ seed
nuclei are photodisintegrated to the iron-peak elements.
Only $N<50$ $p$-nuclides ($^{74}$Se, $^{78}$Kr, and
$^{84}$Sr) are produced in the layers with 
$T_{{\rm m},9}= 3.3$--$3.6$. 
In these hottest layers, the $^{16}$O($\gamma,\alpha$)$^{12}$C
reaction produces $^{12}$C, 
and following $^{12}$C+$^{12}$C fusion reaction produces protons and 
$\alpha$-particles which are identified as the second peak of their
abundances in Figures~\ref{fig3}(a) and \ref{fig3}(c).  
The synthesis of $^{74}$Se, $^{78}$Kr, and $^{84}$Sr through 
proton captures, therefore, becomes effective.  This results in the
increases in the relative yields of the three $p$-nuclides and the
slight increase in the mean average overproduction factor.
However, we find that even at those high temperatures and even for
the lightest three $p$-nuclei, the contribution of proton capture
reactions to the production does not exceed that of
photodisintegration.

\subsubsection{Yields of {\lowercase{p}}-nuclei}

We calculate the overproduction factor $F=X/X_\sun$ which is the 
ratio between the produced abundances and the corresponding solar
abundances for 35 $p$-nuclides in the 21 trajectories.  
The overproduction factors for five $p$-nuclides are plotted as a
function of $T_{\rm m}$ in Figure~\ref{fig4}.  The selected nuclides are
the same as those in Figure~3 of \citet{pra90}.
It is seen from Figure~\ref{fig4} that each nucleus is produced in a
narrow range of the peak temperature.

\placefigure{fig4}

There is one clear difference between SNe~Ia and SNe~II.
The temperature ranges for productions of $p$-nuclei in SNe~Ia are
shifted to hotter layers than in SNe~II \citep{pra90}.  In addition, 
in the W7 model the density of the $p$-process site 
($\sim 10^7\ {\rm g\ cm^{-3}}$) before the explosion 
is higher than in SNe~II ($\sim 10^5\ {\rm g\ cm^{-3}}$).  
Thus, the number of particles (baryons) is larger in SNe~Ia and
the particle-induced reactions are more dominant than in
SNe~II.  It is also the reason why neutron capture reactions,
which become efficient at first in this calculation, drive the nuclear
flow into the neutron-rich region far from $\beta$-stability line
(Section~\ref{sec3}.1.1.).  The 
peak temperature, at which photodisintegration gets predominant, is thus
higher in SNe Ia so that the peak temperature region, in which
$p$-nuclei are created, totally shifts to higher one than in SNe II. 

\begin{figure}
\begin{center}
\includegraphics[angle=-90,scale=0.5]{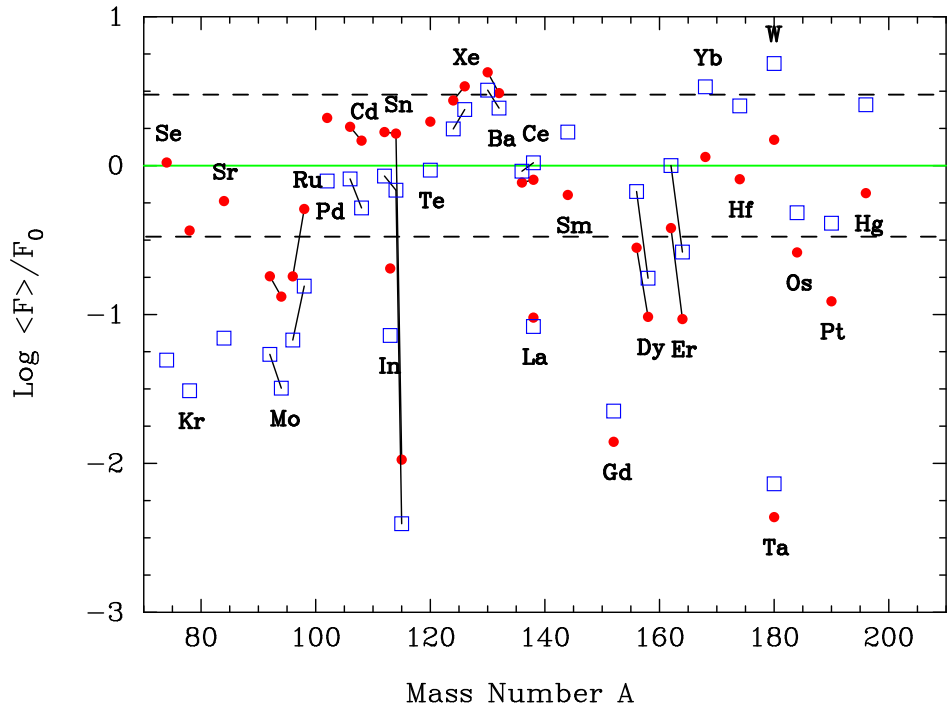}
\end{center}
\caption{Average overproduction factors $\fav$ normalized to $F_0$ for
 the $p$-nuclides as a function of mass number.  The results of Cases
 A1 and B  (Table \ref{tab1}) are shown by filled circles and open squares,
 respectively.  Horizontal solid and dashed lines correspond to $\fav /F_0=1$ and (3, 1/3), respectively.\label{fig5}}
\end{figure}

The overproduction factor averaged over the considered $p$-process
layers is calculated for a $p$-nucleus by the prescription 
\begin{equation}
\fav =\sum^{N}_{j=2}\frac{1}{2}(F_j+F_{j-1})\frac{M_j-M_{j-1}}{M_p},
\label{<F>}
\end{equation}
where $F_j$ is the overproduction factor in a trajectory $j$, $M_j$ is
the Lagrangian mass coordinate of $j$th trajectory, and $M_p$ is the
total mass ($0.137~M_\sun$) of the $p$-process layer.
We define the mean overproduction factor as $F_0=\sum_i  \fav_i/35$, where
$i$ is a species of $p$-nucleus and the summation is taken for $\fav$
values of 35 $p$-nuclei.  
Figure~\ref{fig5} shows the $\fav$ values normalized to $F_0$, in which
the filled circles represent the result for Case~A1.  The numerical
values for these quantities are listed in Table \ref{tab4}.

\placefigure{fig5}
\placetable{tab4}

In Case A1, 19 of 35 $p$-nuclides are produced in
amounts within a factor of three around the mean of $F_0=4657$.  As
mentioned above, the peak temperatures of the used trajectories do not
involve the range of  $1.9 < T_{{\rm m},9} < 2.6$.  Hence, some degrees of
increase in yields are 
expected in a C-deflagration model with finer mesh zoning for nuclei of which 
their production yields increase in the peak temperature region of 
$T_{{\rm m},9}\leq 2.6$ (i.e., $^{138}$La, $^{152}$Gd,
$^{156,158}$Dy, $^{162,164}$Er, $^{168}$Yb, $^{174}$Hf, $^{180}$Ta,
$^{184}$Os, and $^{196}$Hg).  Taking
this fact into account, $^{115}$Sn might be the only $p$-nuclide which is
markedly underproduced.

We compare the result of Case A1 with previous works.  
The pattern of normalized average overproduction factor $\fav/F_0$ for
$p$-nuclides in this calculation is very similar to that in the SNe~II
models~\citep{pra90,ray95}.  For instance, the lightest three nuclides 
($^{74}$Se, $^{78}$Kr, and $^{84}$Sr) are produced more than
the underproduced Mo and Ru isotopes. In the intermediate-mass
$p$-nuclides,  $^{113}$In and 
$^{115}$Sn are also underabundant while the other nuclei show their nearly
equal overproductions.  The most remarkable difference is that in the
C-deflagration model, severe underproductions of the
Mo and Ru $p$-isotopes, which are the most puzzling problem of
the $p$-process in SNe II~\citep{pra90,ray95}, are reduced by a factor
of $\sim 6$--$12$.  However, they still remain 
underproduced with respect to the mean production level for 35 $p$-nuclei.

\subsection{Effect of the Initial Composition of the WD Core on Yields of ${\lowercase{p}}$-nuclides}

\subsubsection{Effect of C/O Ratio}

\begin{figure}
\begin{center}
\includegraphics[angle=-90,scale=0.5]{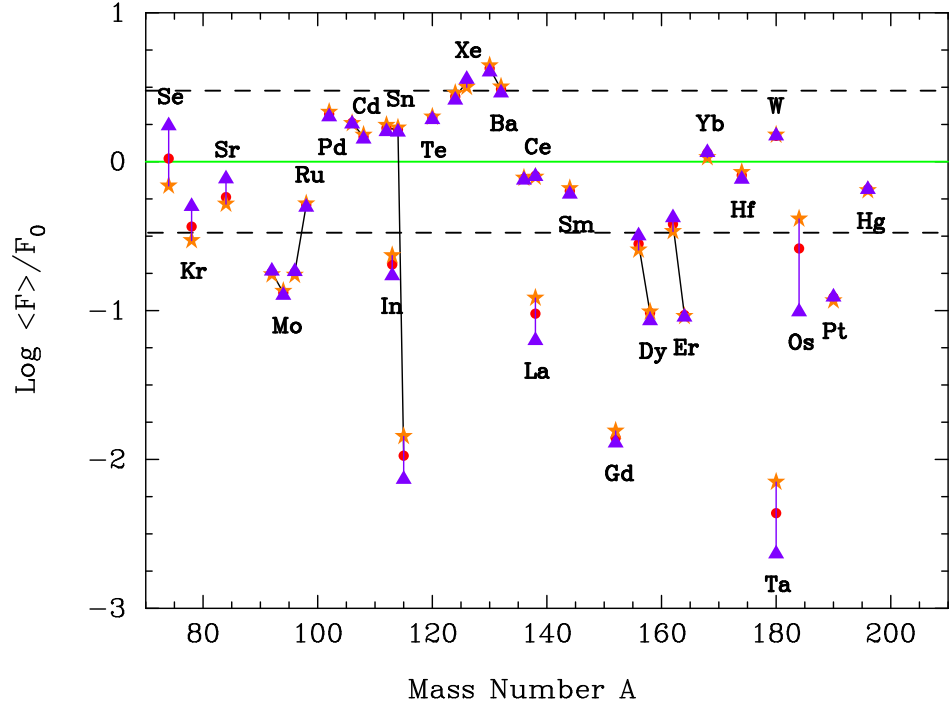}
\end{center}
\caption{Average overproduction factors normalized to $F_0$ for three initial
 compositions: Cases A1 (circles), A2 (stars), A3 (triangles; Table \ref{tab1}).\label{fig6}}
\end{figure}

We investigate the effect of the change in the abundance ratio between
$^{12}$C and $^{16}$O on the $p$-process nucleosynthesis, keeping the
initial abundances of $^{22}$Ne and $s$-nuclei (Case A) fixed.
Figure~\ref{fig6} 
shows the normalized average overproduction factors calculated for
Cases A2 (stars), A1 (circles), and A3 (triangles) in the increasing
order of $^{12}$C/$^{16}$O ratio.  The mean value
$F_0$ for each case is listed in the fifth column of Table~\ref{tab1}.
For the larger C/O ratio, the relative yields of $^{74}$Se, 
$^{78}$Kr, and $^{84}$Sr are larger, and those of $^{115}$Sn,
$^{138}$La, $^{180}$Ta, and $^{184}$Os are smaller, although all the
differences are slight.  There is little variation in the
overproductions $F_0$ (Table~\ref{tab1}).  We, thus, conclude that the
nucleosynthetic results of $p$-process are not so sensitive to the
abundance ratio between $^{12}$C and $^{16}$O if the ratio does not 
affect much the explosion timescale. It should be noted, however, that a variation 
of the C/O ratio could have an influence on the produced amount and
pattern of $p$-nuclei by strongly affecting the explosion timescale.

\placefigure{fig6}

From a comparison of the nuclear flows in Cases A1--A3, we note the
following results.
In layer 5 of Figure~\ref{fig3}, partial C-burning triggers the
production of protons and 
$\alpha$-particles.  The abundance of $\alpha$-particle is, therefore,
higher when the C/O ratio is larger.  Then, the $^{22}$Ne+$\alpha$
reaction proceeds more efficiently, so that the
$^{22}$Ne($\alpha,n$)$^{25}$Mg reaction supplies more neutrons for
larger C/O ratio.

\subsubsection{Effect of $^{22}$Ne Abundance}

Second, we investigate the effect of $^{22}$Ne abundance on the
$p$-process yields by comparing the result of Case A1 with A4. 
In Case A4, the initial abundance of $^{22}$Ne is assumed
to be zero in order to examine the extreme case.

We compare the abundances of light particles in A1 with A4.  The neutron
abundances in A4 are 10--100 times smaller than those in A1 in the
whole range of $p$-process layer.  This is because the neutron supply
through $^{22}$Ne($\alpha,n$)$^{25}$Mg reaction decreases due to no
initial abundance of $^{22}$Ne.  Therefore, the nuclear flow does not
reach so neutron-rich region in A4 at an early phase of the deflagration
wave passage, while a larger amount of neutrons in A1 drive it to a more
neutron-rich region.

\begin{figure}
\begin{center}
\includegraphics[angle=-90,scale=0.5]{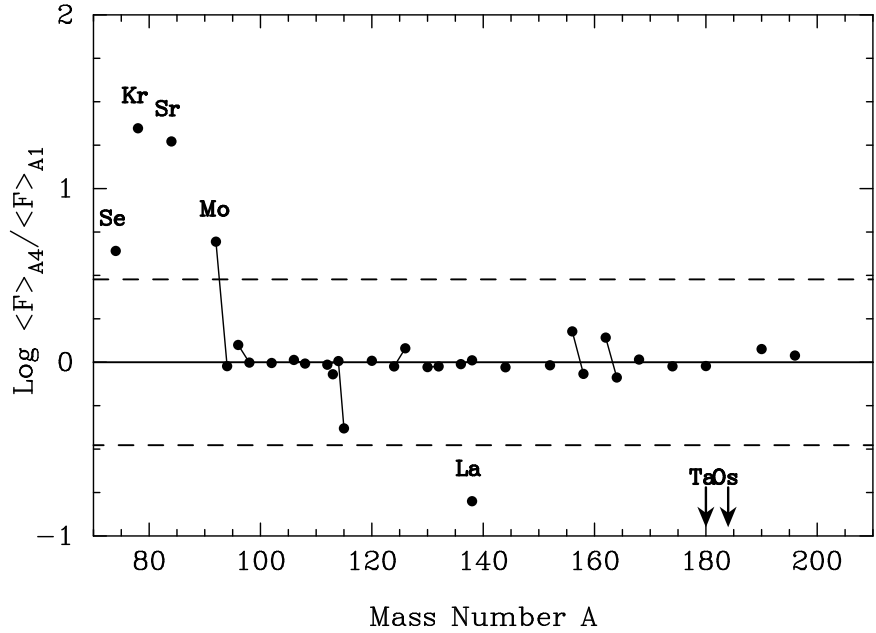}
\end{center}
\caption{Average overproduction factors in Case A4 with respect to those in
 Case A1  (Table \ref{tab1}), i.e., $\fav_{\rm A4}/\fav_{\rm A1}$ for $p$-nucleus.\label{fig7}}
\end{figure}

The ratios of the $\fav$ of each $p$-nuclide in A1 and A4 are plotted in 
Figure~\ref{fig7}. 
It is found that $^{74}$Se, $^{78}$Kr, $^{84}$Sr, and $^{92}$Mo are
produced more abundantly in A4. Since the values of $F_0$ except for 
those four nuclei are almost the same ($\sim 4930$), their efficient
productions are responsible for the increase of $F_0$ in A4. 

\placefigure{fig7}

Next, we investigate the differences of nuclear flow between A1 and A4. 
In the case of A1, $^{74}$Se is created by  
$^{75}$Se($\gamma,n$) reaction in layer 1, but
the ($\gamma,n$), ($\gamma,p$), and ($\gamma,\alpha$) reactions 
photodisintegrate it. Also in layers 2--4, the $^{75}$Se($\gamma,n$)
reaction produces $^{74}$Se, but supplemented by 
$^{73}$As($p,\gamma$) reaction with a small contribution.
For the production of $^{78}$Kr, $^{79}$Kr($\gamma,n$) reaction is
important in layer 1,
but  $^{78}$Kr is photodisintegrated by ($\gamma,p$) reaction. In
layers 2 and 3, $^{78}$Kr is produced by $^{79}$Kr($\gamma,n$) reaction, 
together with a small contribution by $^{77}$Br($p,\gamma$)
reaction. However, in layer 4 $^{77}$Br($p,\gamma$) is a main
production reaction of $^{78}$Kr. For $^{84}$Sr, 
$^{85}$Sr($\gamma,n$) reaction produces it in layers 1--3, but in
layer 1 ($\gamma,n$) and ($\gamma,p$) reactions destruct $^{84}$Sr. In
layer 4, $^{83}$Rb($p,\gamma$) and $^{85}$Sr($\gamma,n$) reactions
enhance the abundance of $^{84}$Sr. $^{92}$Mo is created by the nuclear
flow from $^{93}$Mo($\gamma,n$), 
$^{93}$Tc($\gamma,p$), and $^{96}$Ru($\gamma,\alpha$) reactions, but
the produced $^{92}$Mo is soon destructed by ($\gamma,p$) reaction. 
In layers 2--4, $^{93}$Mo($\gamma,n$) reaction dominates to produce
$^{92}$Mo.

\begin{figure}
\begin{center}
\includegraphics[angle=-90,scale=0.5]{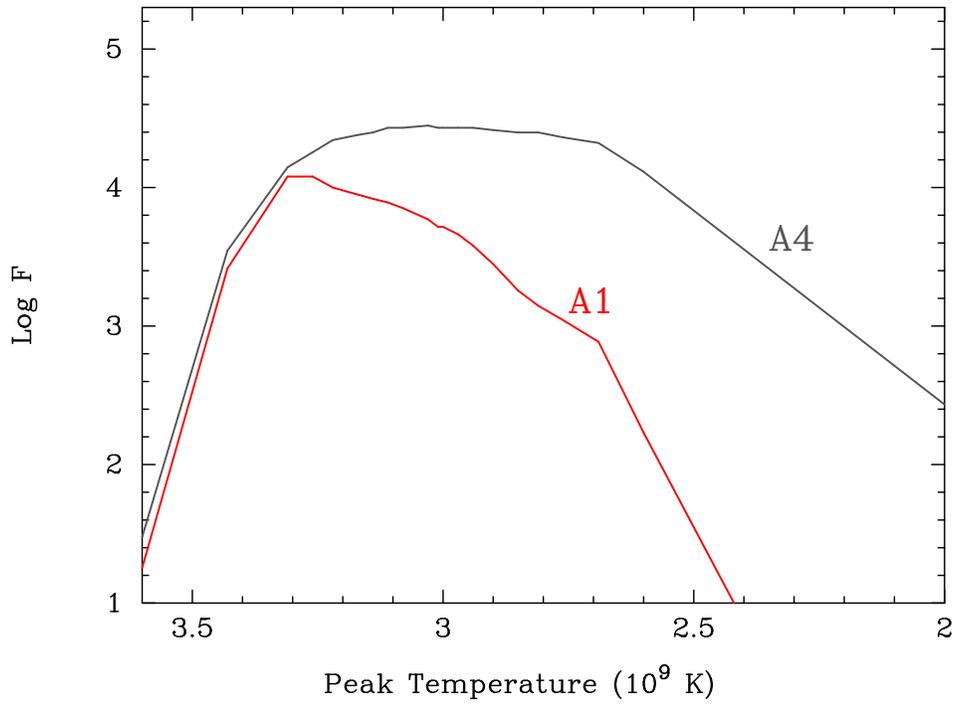}
\end{center}
\caption{Overproduction factors of $^{74}$Se as a function of peak
 temperature for Cases A1 and A4.\label{fig8}}
\end{figure}

In contrast, the proton abundances in A4 are by a factor of $\sim 10$
larger than those in A1. This implies that the larger amount of protons
involves the production of the four $p$-nuclei. In the case of A4, 
$^{73}$As($p,\gamma$) and $^{75}$Se($\gamma,n$) reactions first create
$^{74}$Se in layer 1. After the peak temperature 
exceeds $T_9 = 3$, ($\gamma,p$), ($\gamma,\alpha$) and ($\gamma,n$)
photodisintegrations overcome the production reaction of
$^{75}$Se($\gamma,n$) and destruct $^{74}$Se. In layers 2--4, the
contribution of  
$^{73}$As($p,\gamma$) reaction to the production of $^{74}$Se dominates
that of $^{75}$Se($\gamma,n$) and $^{75}$Br($\gamma,p$) reactions.  
The production of $^{78}$Kr comes from $^{79}$Kr($\gamma,n$)
reaction, together with minor contribution of $^{77}$Br($p,\gamma$)
reaction in layer 1, but the produced $^{78}$Kr is destructed by
($\gamma,p$) reaction. In layers 2--4, the
$^{77}$Br($p,\gamma$) is the dominant reaction to produce $^{78}$Kr,
compared to $^{79}$Kr($\gamma,n$) reaction. For $^{84}$Sr, the production
and destruction processes in layer 1 of A4 are almost the same as in A1.
However, in layers 2--4 $^{83}$Rb($p,\gamma$) reaction has also some
contributions to the production of $^{84}$Sr, together with
$^{85}$Sr($\gamma,n$) reaction. For $^{92}$Mo, the production and
destruction processes in layer 1 of A4 are the same as in A1. In
layers 2--4, $^{91}$Nb($p,\gamma$) reaction is important to produce
$^{92}$Mo and there is a small contribution of $^{93}$Mo($\gamma,n$). This
result is largely different from the case of A1. 
Thus, proton capture reactions have an important role in the production
of lightest four $p$-nuclides. 
Figure~\ref{fig8} shows the overproduction
factor of $^{74}$Se as a function of peak temperature for A1 and A4. 
The important contribution of proton captures to the production of
$^{74}$Se is found in the layer with the peak temperature $T_9 \le 3.3$. 
The production region of almost all $p$-nuclides extends to cooler
layers in A4 than in A1.

\placefigure{fig8}

Figure~\ref{fig7} reveals that $^{180}$Ta and $^{184}$Os are 
underproduced in A4, compared to A1. 
The abundance peak of these two $p$-nuclei in A1 is
found to be in the layer with $T_{\rm m,9} =2.6$.\footnote{The zoning of W7
model in the outermost layer is relatively coarse.  If the finer zones were used, the
abundance peak of $^{180}$Ta and $^{184}$Os would appear in the outer layer
where $T_{\rm m,9}$ would be lower than 2.6.}
In the layer, $^{180}$Ta is first created by $^{181}$Ta($\gamma,n$)
reaction, which is later destructed by $^{180}$Ta($\gamma,n$)
reaction as the temperature increases to its peak. The production and
destruction timescales of $^{180}$Ta depend 
on the inverse reactions. This is because the nuclear flow from
$^{180}$Ta to $^{179}$Ta depends on the rates of $^{180}$Ta($\gamma,n$)
and $^{179}$Ta($n,\gamma$) reactions. 
The reaction rate of photodisintegration is a function of 
temperature only. However, the reaction rate of 
neutron capture is dependent on the temperature, density, and neutron
abundance which is larger in A1 than in A4. 
Thus, the contribution of neutron capture reactions in A1 is more
important than in A4, and the destruction timescale of $^{180}$Ta in A1
is longer than in A4. These facts lead to a
larger abundance of $^{180}$Ta in A1 left after the termination of
nuclear processing. 

The situation for the production of $^{184}$Os is similar to that for
$^{180}$Ta. The production and destruction processes of $^{184}$Os are
the competition between 
photodisintegrations on $^{184}$Os and $^{185}$Os and neutron capture
reactions on $^{183}$Os and $^{184}$Os.
The nuclear flux of
$^{185}$Os($\gamma,n$)$^{184}$Os($\gamma,n$)$^{183}$Os in A1 is
relatively small due to a larger amount of neutrons than in A4. In 
addition, $^{183}$Os($n,\gamma$) creates $^{184}$Os in the later phase
of explosion in A1. As a result, the final abundance of $^{184}$Os is
by a factor of 15 larger in A1. 


\subsection{Effect of the Initial Abundances of ${\lowercase{s}}$-nuclei
  on Yields of ${\lowercase{p}}$-nuclei}

Here, we examine the impact of different seed distributions of Cases A
(A1--A4) and B with the enhancement from the solar abundance of heavy
nuclei.  In this investigation, we compare A1 with B, which has
similar abundances to A1 for $^{12}$C, $^{16}$O,
and $^{22}$Ne in the CO core.

Figure~\ref{fig2} shows that the overabundances of $s$-nuclei in Case B
are by factors of 10--100 larger than those in A. Especially, in B heavy
seed $s$-nuclei with $A>140$ are about 100 times more abundant than in
A. This leads to the increase of $F_0$ by a factor of 50 in B, relative
to A. This increase is seen for each $p$-nuclide in Figure~\ref{fig5}, 
in which the normalized
$\fav$ is presented for Cases A1 and B. The values of normalized $\fav$
for heavy $p$-nuclei with $A>140$, especially $^{168}$Yb, $^{174}$Hf,
$^{180}$W, and $^{196}$Hg, are larger than those in A. In contrast,
lighter $p$-nuclei have smaller values in B than those in A. In
particular, $^{74}$Se--$^{98}$Ru shows significant reductions, which come
from the decrease in relative abundances of seed nuclei near
the $p$-nuclei, compared to heavy seed nuclei with $A>140$.

The resulting distribution of $p$-nuclei shows important decreases in
$^{74}$Se--$^{98}$Ru as seen in B, if the distribution of heavy seed
$s$-nuclei is 
similar to that of the solar distribution. This result is analogous 
to that obtained by the $p$-process in SNe~II, but with the
even worse reduction of $^{74}$Se--$^{84}$Sr. Thus, we find that the
distribution similar to that of the solar $p$-nuclei is obtained
by the seed distribution attributed to the $s$-process under
an environment with metallicity close to solar~\citep{gal98}, although
main neutron source might be $^{22}$Ne($\alpha,n$) reaction. 

The presupernova abundance of $s$-nuclei in the WD 
thus has a large impact on the $p$-process nucleosynthesis and
is very crucial for the resulting yields of $p$-nuclei.

\subsection{Contribution of C-deflagration SNe to the Galactic Yields of
  ${\lowercase{p}}$-nuclei} 

The biggest question concerning the $p$-process nucleosynthesis is where
the main production site is.  
In order to investigate this question, we estimate the contribution of 
SN~Ia to galactic chemical evolution of $p$-nuclei by comparing the
ejected mass of $^{56}$Fe, a main product of SN~Ia, with those of
$p$-nuclei in Case A1. 

Here, we introduce a {\it net yield} of each nuclide in SNe~Ia,
defined as a difference between the mass of the nuclide returned to 
the interstellar space at the SN explosion and the mass engulfed into
a star at its birth. 
The $p$-nuclei present at the star formation remain inside the 
$p$-process layer before the explosion, but they are destroyed 
by photodisintegrations. In addition, 
we assume that the $p$-nuclei present in an accreted matter are 
burned by neutron capture during the $s$-process developing in the He
intershell (see, e.g., Tables~\ref{tab6} and \ref{tab7}).  
Therefore, the $p$-nuclei ejected from the SN~Ia explosion are only those 
ultimately produced in the $p$-process layer.

Assuming that the mass of
CO core in the WD is approximately equal to that of the whole
WD, the net yield for $p$-nuclide $i$ is given by
\begin{equation}
y_i=X_{i,\sun} (\fav_{i}M_{\it p}-M_{\rm WD}),
\label{yi}
\end{equation}
where $M_{\rm WD}=1.378M_{\sun}$ is the mass of the WD \citep{nom84} and
$X_{i,\sun}$ is the solar mass fraction of $p$-nuclide $i$.  
The above prescription could also be applied to $^{56}$Fe, so that the
net yield of $^{56}$Fe is
\begin{equation}
y_{^{56}{\rm Fe}}=X_{^{56}{\rm Fe},\sun}
   (M_{^{56}{\rm Fe}}/X_{^{56}{\rm Fe},\sun}-M_{\rm WD}), 
\label{yfe}
\end{equation}
where $M_{^{56}{\rm Fe}}$ is the mass of $^{56}$Fe ejected by an SN~Ia
and $X_{^{56}{\rm Fe},\sun} = 1.17\times10^{-3}$ is the solar mass
fraction of $^{56}$Fe.  By taking a ratio between these quantities 
normalized to the corresponding solar mass fraction for respective
nuclides, we can evaluate how many SN~Ia events contribute to the
galactic chemical evolution of one nuclide.  The calculated ratio of
yields between $^{56}$Fe and a representative $p$-nucleus is as follows:
\begin{eqnarray}
^{56}{\rm Fe}/p&\equiv&\frac{y_{^{56}{\rm Fe}}/X_{^{56}{\rm
 Fe},\sun}}{(1/35)\sum^{35}_{i=1} y_{i}/X_{i,\sun}} \nonumber \\
&=&\frac{M_{^{56}{\rm
 Fe}}/X_{^{56}{\rm Fe},\sun}-M_{\rm WD}}{F_0 M_p-M_{\rm WD}}
\label{ratio}
\end{eqnarray}
where the yield of the $p$-nucleus is derived by summing the
net yields of each $p$-nuclide and averaging it over 35 $p$-nuclei.

The yield ratio ($^{16}$O/{\it p}) between $^{16}$O and the representative 
$p$-nucleus is also defined in the same way, in which the solar mass
fraction of $^{16}$O is assumed to be 
$X_{^{16}{\rm O},\sun}= 9.59\times10^{-3}$.
In this C-deflagration model for an SN~Ia, we obtain $^{56}$Fe/{\it p}
$=0.82$ and $^{16}$O/{\it p} $=0.023$ for Case A1~\citep{tsu95}.
These results indicate that $p$-nuclei are produced more than $^{56}$Fe
and $^{16}$O when normalized to their solar
abundances.  We, therefore, conclude that SNe~Ia can account for the
galactic content of $p$-nuclei assuming that the seed $s$-nuclei are
produced efficiently up to the overproduction level of 
$\sim 10^3$ in He-rich layer accreting onto the degenerate CO WD.

\subsection{Comparison with Previous ${\lowercase{p}}$-process Calculations}

\begin{figure}
\begin{center}
\includegraphics[angle=-90,scale=0.5]{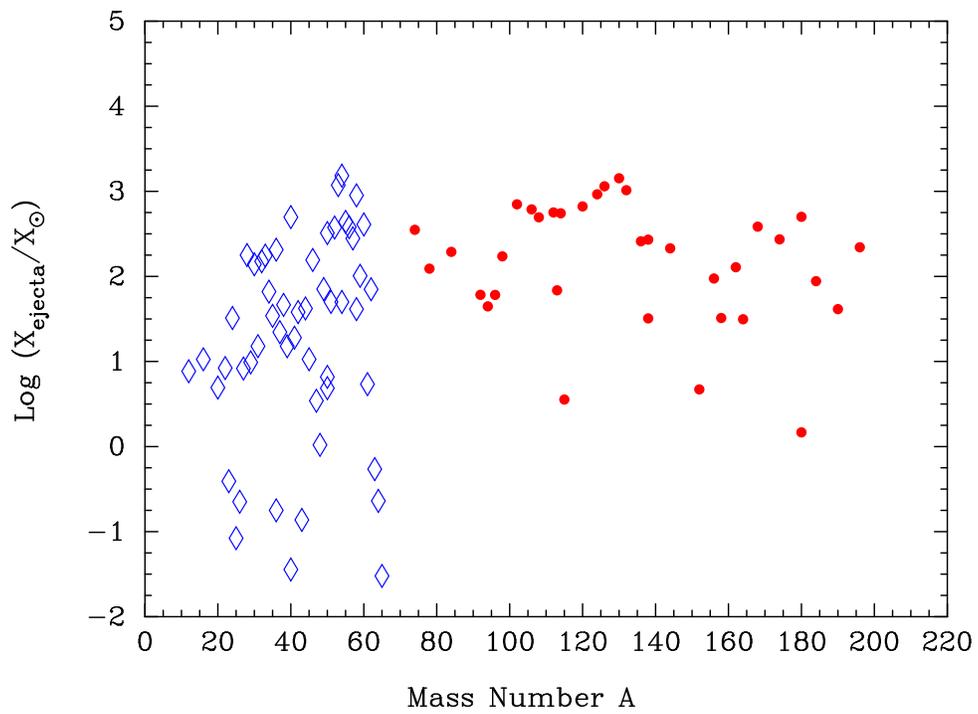}
\end{center}
\caption{Overproduction of $p$-nuclei (filled circles) and nuclei
 lighter than Fe-group elements (open diamonds) in the ejecta of SNe~Ia.\label{fig9}}
\end{figure}

\citet{gor02} calculated $p$-process nucleosynthesis in the
sub-Chandrasekhar mass He-detonation model.  The 
stable nuclei from Ca to Fe are overabundant with respect to 
$p$-nuclei by a factor of $\sim 100$. They, therefore, concluded that
the He-detonation model is not an efficient site for production of
$p$-nuclei.   
Nevertheless, \citeauthor{gor02} claimed that if the initial
abundances of $s$-nuclei are enhanced over their solar values by a
factor of 100, 
$p$-nuclei are produced at the same level as Ca--Fe, while
such enhancement of the $s$-nuclei is not trivial in the
He-detonation model.  
On the other hand, in the present C-deflagration model
the overproduction factors for ejected $p$-nuclides and stable
nuclides lighter than Fe-group elements are plotted as a function of mass
number in Figure~\ref{fig9} and are listed in Table~\ref{tab5}.
From this figure, it is found that $F_{\rm Ca\mathchar` Fe} \sim F_p$ is realized
with some enhancements of heavy seed nuclei, which are a natural 
consequence from the accretion of H-rich matter onto a CO WD with an
appropriate rate of mass accretion, 
although the enhancement level of heavy seed nuclei remains ambiguous. 
This means that the C-deflagration model is $\sim$100 times more
effective in enriching a galaxy with $p$-nuclei, if the
assumed level of enhancement is valid.

\placefigure{fig9}
\placetable{tab5}

As mentioned in Section 1, the DD model for SNe~Ia may
also produce an important amount of $p$-nuclei if the prior enhancement
of $s$-nuclei in a Chandrasekhar mass WD model is taken into account
\citep{how92} as in our present assumptions.  
However, the enhancement of $s$-nuclei is not trivial for the double
degenerate scenario of SN~Ia in a Chandrasekhar mass model.

\citet{arn03} have calculated the $p$-process nucleosynthesis in the
W7 model for SNe Ia adopting two cases as an initial abundance of heavy
seed nuclei. The resulting pattern of yields of $p$-nuclei in
\citet{arn03}  under the assumption of the solar abundance for initial
seed shows the following difference from our Case A. 
Normalized abundances of $p$-nuclei for Mo--Ce in \citet{arn03} are much
smaller, but those for heavier $p$-nuclei are larger. This trend 
represents the solar abundance of $p$-nuclei themselves. On the other hand,
their result for the  
initial seed distribution representative of the $s$-process in AGB stars
of the solar metallicity is similar to that of Case A.  It is, however,
found that there exists a difference of the pattern between 
our Case A and \citet{arn03} for $p$-nuclei with $N> 82$, while the pattern
for $p$-nuclei with $N \le 82$ is similar. This result represents
that the abundance pattern for heavier $p$-nuclei is very sensitive to
initial seed distributions. It should be noted that the
differences in yields of $^{113}$In, $^{158}$Dy, and $^{190}$Pt are
larger than those within nuclear uncertainties evaluated by
\citet{arn03}.

\citet{pra90} studied the $p$-process in SNe~II using the model for
SN1987A, and obtained $F_{\rm O}/F_p$ ($\approx ^{16}$O/{\it p}) $\sim
12$, where $F_{\rm O}$ stands 
for the overproduction factor of oxygen. They claimed that the solar
system content of $p$-nuclei does not come from SN1987A-like SNe. One
reason may be attributed to the abundances of 
seed nuclei which reflect low metallicity in the Large Magellanic Cloud. 
\citet{ray95} derived the ratios of yields for oxygen and $p$-nuclei from 
star models with masses of 13, 15, 20, and 25~$M_\sun$ and averaged them
over the initial mass function (IMF).  They found that the resulting
ratio was $F_{\rm O}/F_p =4.2$.  
These results, thus, imply that $p$-nuclei are not sufficiently produced,
compared to the main product in SNe~II and suggest that the other
production sites have an 
important contribution to the galactic evolution of $p$-nuclei. 
Accordingly, the C-deflagration model for SNe~Ia is one of the
promising sites for the $p$-process because sufficient amounts of
$p$-nuclei are produced relative to $^{16}$O within our assumptions.

Using the results of this study and previous works, we estimate how many
SNe~Ia and SNe~II contribute to the galactic evolution
of $p$-nuclei.  We assume that all SNe~Ia would typically
experience the $p$-process as in Case A1.  First, we adopt the
ratios of overproduction factors $^{56}$Fe/{\it p} $=0.82$ for our SN~Ia
model  and $^{16}$O/{\it p} $=4.2$ for SNe~II \citep{ray95}.
We also take the ejected masses of $^{16}$O and $^{56}$Fe from the W7
model for SN~Ia and the IMF-averaged masses of these nuclides from
SNe~II \citep{tsu95}.  These masses normalized to 
the corresponding solar abundances are summarized in
Table~\ref{tab2}.  Values for $p$-nuclei are estimated from the yield
ratios between $p$-nuclei and the main products ($^{56}$Fe for SN~Ia
and $^{16}$O for SN~II).
In order to compare these yields between SN~Ia and SN~II,
we adopt the ratio of the occurrence frequency of SNe~Ia to that of
SNe~II, $\sim 0.15$ \citep{tsu95}.  From the values in Table~\ref{tab3},
we find that SNe~Ia contribute about 70\% of the 
$p$-nuclei in the solar system.  It should be noted that there is
still a small underproduction of $p$-nuclei with respect to $^{16}$O,
i.e., $^{16}$O/{\it p} $=2/1.46=1.4$, while the ratio of the yields between
$^{56}$Fe and $p$-nuclei almost agrees with that of the solar abundance,
i.e., $^{56}$Fe/{\it p} $=1.57/1.46=1.1$. 

\placetable{tab2}
\placetable{tab3}

\section{CONCLUSIONS}\label{sec4}
We calculate the $p$-process nucleosynthesis in the carbon-deflagration
model for SNe~Ia (W7 model in Nomoto et al. 1984) with realistic initial 
abundances of $s$-nuclei.  The temperature and density
trajectories in W7 are different from those of the DD model adopted in
\cite{how92} and the parameter study~\citep{how91}.  The adopted
$s$-process patterns are also different from the previous study.  We
investigate the effects on productions of $p$-nuclei of (1) 
the different initial $^{12}$C abundances which affect the explosive
C-burning, (2) the uncertain initial abundance of $^{22}$Ne 
at the explosion, whose ($\alpha,n$) reaction is an important neutron 
source for the $s$-process occurring during the He shell flashes, and 
(3) the different distributions in the $s$-process which provides
initial seed abundances for the $p$-process.  Our findings are
summarized as follows. 

1. In all cases we considered, more than 50\% of $p$-nuclides are
co-produced at almost the same degree of enhancements with respect to
their solar abundances.  We
find that SNe~Ia can produce Mo and Ru $p$-isotopes $\sim 6$--$12$ times
more in Case A1 than SNe~II on the basis of the mean overproduction
factor of $p$-nuclei, although the problem of the relative
underproduction still remains.  
The patterns of $p$-nuclei obtained in this study are
different from those in~\citet{how92} and~\cite{how91}, reflecting the
differences in temperature and density profiles and initial
distribution of seed nuclei.  
In addition, it is confirmed that the $p$-process layer in this
C-deflagration model shifts to a higher temperature region than that in
SNe~II. 

2. The effect of variable C/O ratio in the initial composition of the CO
WD on the $p$-nuclei yields is small.  On the other hand, the effect of
the initial $^{22}$Ne abundances is large especially for $^{74}$Se,
$^{78}$Kr, $^{84}$Sr, and $^{92}$Mo.  If the initial $^{22}$Ne is less
abundant, the light $p$-nuclides are enhanced.  In order to produce
enough Mo and Ru $p$-isotopes, the abundance of $^{22}$Ne smaller than
the assumption in W7 model is expected.  Initial 
abundances of various nuclides other than $^{12}$C, $^{16}$O, and
$^{22}$Ne considered in this study also affect the $p$-process through
supplies of neutron, proton, and $\alpha$-particle emerging during the
explosion.

3. The effect of initial abundances of $s$-nuclei on the $p$-process is
large. If the $s$-process efficiently contributes to the production of
seed nuclei, yields of $p$-nuclei increase and relative yields of heavy
$p$-nuclei are enhanced more than those of lighter ones.  
Such enhancement of $s$-nuclei is expected for the single degenerate
scenario of the Chandrasekhar mass model, but not trivial in the
double degenerate scenario.

4. This result leads to a possibility that SNe~Ia play an important role
in the galactic evolution of $p$-nuclei. The $p$-nuclei are produced by
a factor of $\sim 1.2$ more than $^{56}$Fe when normalized to the solar
abundances. SNe~Ia, therefore, may have contributed to the enrichment of
$p$-nuclei more effectively than SNe~II (about twice in Case A1).
Our calculation involves an uncertainty in the initial abundances of
nuclides (e.g., $^{22}$Ne) in the C-deflagration model of the exploding
CO WD. This has a great influence on abundances of background 
particles, i.e., protons, neutrons, and $\alpha$-particles in the
$p$-process nucleosynthesis.  
There is also an important uncertainty in the $s$-process
in presupernova stars which affects initial distributions of the
$s$-nuclei and the remaining $^{22}$Ne abundance.  Studies of the
$s$-process nucleosynthesis during the presupernova evolution of accreting
WDs are necessary to clarify if the $p$-process in SNe~Ia
could really produce enough abundances of $p$-nuclei.  Despite these
uncertainties,  the present study strongly suggests that SNe~Ia,
as well as SNe~II, could be a probable production site of the solar $p$-nuclei.

\acknowledgments
This work has been supported by Grant-in-Aid for the Japan Society for
the Promotion of Science (JSPS) Fellows (21.6817), and for
Scientific Research of JSPS (18104003, 18540231, 20244035, 20540226) and
the Ministry of Education, Culture, Sports, Science, and Technology
(MEXT; 19047004, 20040004).  This work has also been supported by World
Premier International Research Center Initiative, MEXT, of Japan.

\clearpage

\clearpage

\begin{deluxetable}{cc}
\tablecaption{Initial Mass Fractions $X$ of Seed Nuclei for Case A
 \label{tab6}}
\tablewidth{0pt}
\tablehead{
\colhead{Nuclide} & \colhead{Abundance}}
 \startdata
  $^{32}$S &  1.92E-04   \\
  $^{33}$S &  1.03E-04   \\
  $^{34}$S &  7.35E-05   \\
  $^{36}$S &  4.99E-05   \\
  $^{35}$Cl &  3.79E-05
\enddata
\tablecomments{(This table is available in its entirety in a machine-readable form in the online journal. A
portion is shown here for guidance regarding its form and content.)}
\end{deluxetable}

\begin{deluxetable}{cc}
\tablecaption{Initial Mass Fractions $X$ of Seed Nuclei for Case B
 \label{tab7}}
\tablewidth{0pt}
\tablehead{
\colhead{Nuclide} & \colhead{Abundance}}
\startdata
  $^{32}$S &  1.35E-04  \\
  $^{33}$S &  8.90E-05  \\
  $^{34}$S &  9.04E-05  \\
  $^{36}$S &  1.85E-04  \\
 $^{35}$Cl &  6.22E-05
\enddata
\tablecomments{(This table is available in its entirety in a machine-readable form in the online journal. A
portion is shown here for guidance regarding its form and content.)}
\end{deluxetable}
 

\begin{deluxetable}{ccccc}
\tablecaption{Initial Compositions and Average Overproduction Factors of
 $p$-nuclei\label{tab1}}
\tablewidth{0pt}
\tablehead{
\colhead{Case} & \colhead{$X$($^{12}$C)} & \colhead{$X$($^{16}$O)} &
 \colhead{$X$($^{22}$Ne)} & \colhead{$F_0$}
}
\startdata
A1& 0.475 & 0.500 & 0.023 & 4657 \\
A2& 0.350 & 0.625 & 0.023 & 4606 \\
A3& 0.700 & 0.275 & 0.023 & 4777 \\
A4& 0.498 & 0.500 & 0.000 & 7621 \\
B & 0.475 & 0.500 & 0.022 & 250734
\enddata

\end{deluxetable}

\clearpage

\begin{deluxetable}{cccccc}
\tablecaption{Average Overproduction Factors $\fav$ Normalized to $F_0$
 for $p$-nuclei \label{tab4}}
\tablewidth{0pt}
\tablehead{
\colhead{Nuclide} & \colhead{Case A1} & \colhead{Case A2} &
 \colhead{Case A3} & \colhead{Case A4} & \colhead{Case B}
}
\startdata
  $^{74}$Se &  1.05E+00 &  6.87E-01 &  1.74E+00 &  2.81E+00 &  4.94E-02 \\
  $^{78}$Kr &  3.67E-01 &  2.96E-01 &  5.01E-01 &  4.98E+00 &  3.07E-02 \\
  $^{84}$Sr &  5.78E-01 &  5.19E-01 &  7.68E-01 &  6.59E+00 &  6.94E-02 \\
  $^{92}$Mo &  1.81E-01 &  1.74E-01 &  1.84E-01 &  5.46E-01 &  5.40E-02 \\
  $^{94}$Mo &  1.32E-01 &  1.35E-01 &  1.27E-01 &  7.66E-02 &  3.20E-02 \\
  $^{96}$Ru &  1.80E-01 &  1.73E-01 &  1.83E-01 &  1.39E-01 &  6.74E-02 \\
  $^{98}$Ru &  5.12E-01 &  5.21E-01 &  4.95E-01 &  3.11E-01 &  1.55E-01 \\
 $^{102}$Pd &  2.09E+00 &  2.15E+00 &  2.00E+00 &  1.26E+00 &  7.89E-01 \\
 $^{106}$Cd &  1.82E+00 &  1.81E+00 &  1.79E+00 &  1.15E+00 &  8.15E-01 \\
 $^{108}$Cd &  1.47E+00 &  1.51E+00 &  1.42E+00 &  8.83E-01 &  5.20E-01 \\
 $^{113}$In &  2.04E-01 &  2.34E-01 &  1.71E-01 &  1.06E-01 &  7.23E-02 \\
 $^{112}$Sn &  1.68E+00 &  1.76E+00 &  1.59E+00 &  9.94E-01 &  8.51E-01 \\
 $^{114}$Sn &  1.64E+00 &  1.68E+00 &  1.58E+00 &  1.02E+00 &  6.84E-01 \\
 $^{115}$Sn &  1.06E-02 &  1.43E-02 &  7.36E-03 &  2.70E-03 &  3.93E-03 \\
 $^{120}$Te &  1.98E+00 &  1.99E+00 &  1.92E+00 &  1.23E+00 &  9.32E-01 \\
 $^{124}$Xe &  2.74E+00 &  2.88E+00 &  2.59E+00 &  1.58E+00 &  1.76E+00 \\
 $^{126}$Xe &  3.41E+00 &  3.16E+00 &  3.58E+00 &  2.51E+00 &  2.38E+00 \\
 $^{130}$Ba &  4.23E+00 &  4.41E+00 &  4.01E+00 &  2.43E+00 &  3.21E+00 \\
 $^{132}$Ba &  3.07E+00 &  3.17E+00 &  2.90E+00 &  1.77E+00 &  2.43E+00 \\
 $^{138}$La &  9.54E-02 &  1.21E-01 &  6.30E-02 &  9.24E-03 &  8.31E-02 \\
 $^{136}$Ce &  7.69E-01 &  7.75E-01 &  7.53E-01 &  4.59E-01 &  9.18E-01 \\
 $^{138}$Ce &  8.03E-01 &  7.87E-01 &  7.96E-01 &  5.03E-01 &  1.04E+00 \\
 $^{144}$Sm &  6.36E-01 &  6.59E-01 &  6.06E-01 &  3.63E-01 &  1.68E+00 \\
 $^{152}$Gd &  1.40E-02 &  1.55E-02 &  1.29E-02 &  8.20E-03 &  2.25E-02 \\
 $^{156}$Dy &  2.81E-01 &  2.55E-01 &  3.18E-01 &  2.58E-01 &  6.70E-01 \\
 $^{158}$Dy &  9.64E-02 &  9.79E-02 &  8.55E-02 &  5.05E-02 &  1.75E-01 \\
 $^{162}$Er &  3.81E-01 &  3.40E-01 &  4.21E-01 &  3.22E-01 &  1.00E+00 \\
 $^{164}$Er &  9.33E-02 &  9.14E-02 &  9.05E-02 &  4.65E-02 &  2.63E-01 \\
 $^{168}$Yb &  1.14E+00 &  1.07E+00 &  1.15E+00 &  7.23E-01 &  3.38E+00 \\
 $^{174}$Hf &  8.10E-01 &  8.46E-01 &  7.63E-01 &  4.69E-01 &  2.52E+00 \\
 $^{180}$Ta &  4.36E-03 &  7.04E-03 &  2.32E-03 &  2.06E-04 &  7.29E-03 \\
 $^{180}$W  &  1.49E+00 &  1.51E+00 &  1.49E+00 &  8.67E-01 &  4.85E+00 \\
 $^{184}$Os &  2.61E-01 &  4.12E-01 &  9.82E-02 &  1.19E-02 &  4.83E-01 \\
 $^{190}$Pt &  1.23E-01 &  1.16E-01 &  1.23E-01 &  8.92E-02 &  4.11E-01 \\
 $^{196}$Hg &  6.54E-01 &  6.39E-01 &  6.51E-01 &  4.36E-01 &  2.57E+00 
\enddata
\tablecomments{All values in respective columns are normalized to
 different $F_0$ values, which are listed in Table~\ref{tab1}.}
\end{deluxetable}

\clearpage

\begin{deluxetable}{cccc|cc}
\tablecaption{Overproduction Factors of Ejecta $X_{\rm ejecta}/X_\sun$
 for Stable Nuclei from C to Fe and $p$-nuclei \label{tab5}}
\tablewidth{0pt}
\tablehead{
\colhead{Nuclide} & \colhead{Abundance} & \colhead{Nuclide} &
 \colhead{Abundance} & \colhead{Nuclide} & \colhead{Abundance}
}
\startdata
   $^{12}$C &  7.66E+00 &   $^{46}$Ca &  1.79E-03 &  $^{74}$Se & 3.53E+02 \\
   $^{13}$C &  1.51E-08 &   $^{48}$Ca &  1.15E-06 &  $^{78}$Kr & 1.23E+02 \\
   $^{14}$N &  3.15E-06 &   $^{45}$Sc &  1.06E+01 &  $^{84}$Sr & 1.94E+02 \\
   $^{15}$N &  3.49E-03 &   $^{46}$Ti &  1.56E+02 &  $^{92}$Mo & 6.07E+01 \\
   $^{16}$O &  1.06E+01 &   $^{47}$Ti &  3.45E+00 &  $^{94}$Mo & 4.44E+01 \\
   $^{17}$O &  5.42E-05 &   $^{48}$Ti &  1.05E+00 &  $^{96}$Ru & 6.06E+01 \\
   $^{18}$O &  1.51E-07 &   $^{49}$Ti &  7.10E+01 &  $^{98}$Ru & 1.72E+02 \\
   $^{19}$F &  5.20E-05 &   $^{50}$Ti &  4.86E+00 & $^{102}$Pd & 7.03E+02 \\
  $^{20}$Ne &  4.93E+00 &    $^{50}$V &  6.59E+00 & $^{106}$Cd & 6.13E+02 \\
  $^{21}$Ne &  7.21E-03 &    $^{51}$V &  5.01E+01 & $^{108}$Cd & 4.94E+02 \\
  $^{22}$Ne &  8.36E+00 &   $^{50}$Cr &  3.23E+02 & $^{113}$In & 6.85E+01 \\
  $^{23}$Na &  3.91E-01 &   $^{52}$Cr &  3.76E+02 & $^{112}$Sn & 5.65E+02 \\
  $^{24}$Mg &  3.24E+01 &   $^{53}$Cr &  1.18E+03 & $^{114}$Sn & 5.52E+02 \\
  $^{25}$Mg &  8.37E-02 &   $^{54}$Cr &  5.00E+01 & $^{115}$Sn & 3.57E+00 \\
  $^{26}$Mg &  2.24E-01 &   $^{55}$Mn &  4.37E+02 & $^{120}$Te & 6.64E+02 \\
  $^{27}$Al &  8.26E+00 &   $^{54}$Fe &  1.53E+03 & $^{124}$Xe & 9.21E+02 \\
  $^{28}$Si &  1.78E+02 &   $^{56}$Fe &  3.79E+02 & $^{126}$Xe & 1.15E+03 \\
  $^{29}$Si &  9.75E+00 &   $^{57}$Fe &  2.80E+02 & $^{130}$Ba & 1.42E+03 \\
  $^{30}$Si &  1.39E+02 &   $^{58}$Fe &  4.12E+01 & $^{132}$Ba & 1.03E+03 \\
   $^{31}$P &  1.51E+01 &   $^{59}$Co &  1.02E+02 & $^{138}$La & 3.21E+01 \\
   $^{32}$S &  1.50E+02 &   $^{58}$Ni &  8.96E+02 & $^{136}$Ce & 2.59E+02 \\
   $^{33}$S &  1.76E+02 &   $^{60}$Ni &  4.08E+02 & $^{138}$Ce & 2.70E+02 \\
   $^{34}$S &  6.61E+01 &   $^{61}$Ni &  5.41E+00 & $^{144}$Sm & 2.14E+02 \\
   $^{36}$S &  1.78E-01 &   $^{62}$Ni &  7.06E+01 & $^{152}$Gd & 4.70E+00 \\
  $^{35}$Cl &  3.44E+01 &   $^{64}$Ni &  2.30E-01 & $^{156}$Dy & 9.44E+01 \\
  $^{37}$Cl &  2.21E+01 &   $^{63}$Cu &  5.43E-01 & $^{158}$Dy & 3.24E+01 \\
  $^{36}$Ar &  2.06E+02 &   $^{65}$Cu &  3.02E-02 & $^{162}$Er & 1.28E+02 \\
  $^{38}$Ar &  4.63E+01 &             &           & $^{164}$Er & 3.14E+01 \\
  $^{40}$Ar &  3.59E-02 &             &           & $^{168}$Yb & 3.84E+02 \\
   $^{39}$K &  1.51E+01 &             &           & $^{174}$Hf & 2.72E+02 \\
   $^{41}$K &  1.90E+01 &             &           & $^{180}$Ta & 1.47E+00 \\
  $^{40}$Ca &  4.97E+02 &             &           & $^{180}$W  & 5.01E+02 \\
  $^{42}$Ca &  3.81E+01 &             &           & $^{184}$Os & 8.78E+01 \\
  $^{43}$Ca &  1.38E-01 &             &           & $^{190}$Pt & 4.12E+01 \\
  $^{44}$Ca &  4.19E+01 &             &           & $^{196}$Hg & 2.20E+02
\enddata
\end{deluxetable}

\clearpage

\begin{deluxetable}{cccc}
\tablecaption{Relative Ejected Mass in One SN Event \label{tab2}}
\tablewidth{0pt}
\tablehead{
\colhead{Type} & \colhead{$p$-nuclei\tablenotemark{a}} & \colhead{$^{16}$O\tablenotemark{a}} & \colhead{$^{56}$Fe\tablenotemark{a}}
}
\startdata
SN II& 45 & 188 & 72 \\
SN Ia& 638 & 15 & 524 
\enddata
\tablenotetext{a}{Values of $M_i/X_{i,\sun}$ for nuclei $i$ and average
 over $p$-nuclei.}
\tablecomments{As for values of SNe II, the IMF averages in \citealt{tsu95} are used.}

\end{deluxetable}



\begin{deluxetable}{cccc}
\tablecaption{Contributions to the Galactic Contents of $p$-nuclei \label{tab3}}
\tablewidth{0pt}
\tablehead{
\colhead{Type} & \colhead{$p$-nuclei} & \colhead{$^{16}$O} & \colhead{$^{56}$Fe}
}
\startdata
SN II& 0.46 & 2.0 & 0.75 \\
SN Ia& 1 & 0 & 0.82 
\enddata
\tablecomments{Ratios of masses that have ever been ejected in the
 Galaxy by two types of SN events to the corresponding solar mass
 fractions.  Values are normalized so that the $p$-nuclei
 for SNe Ia are unity.}

\end{deluxetable}

\end{document}